%% file: main.tex
\documentclass[sigplan,nonacm]{acmart}
\renewcommand\footnotetextcopyrightpermission[1]{}
\usepackage{algorithm}
\usepackage{algpseudocode}
\usepackage{amsmath}
\usepackage{graphicx}
\usepackage{lipsum}
\usepackage{xspace}
\usepackage{soul}
\usepackage{enumitem}
\usepackage{my-latex-commands}

\newcommand{\sysname}{OrbitWeaver\xspace}
\newcommand{\satelliteone}{Sat-1\xspace}
\newcommand{\satellitetwo}{Sat-2\xspace}
\newcommand{\para}[1]{\noindent\textbf{#1}\xspace}
\newcommand{\parabullet}[1]{\noindent$\bullet$ \textbf{#1}\xspace}

\begin{document}
    \setlength{\abovedisplayshortskip}{1pt}
    \setlength{\belowdisplayshortskip}{1pt}
    \setlength\abovecaptionskip{2pt}
    \setlength\belowcaptionskip{2pt}
    \setlength{\textfloatsep}{2pt}
    \setlength{\dbltextfloatsep}{2pt}
    \setlength{\floatsep}{2pt}
    \setlength{\dblfloatsep}{2pt}
    \setlength{\intextsep}{2pt}

    \title{Resuming Vision-Language Inference across Onboard Interruptions}
    \author{Xiaolong Huang}
    \affiliation{\institution{Peking University}\city{Beijing}\country{China}}
    \email{huangxiaolong@pku.edu.cn}
    \author{Yinyuan Zhang}
    \affiliation{\institution{Peking University}\city{Beijing}\country{China}}
    \email{yinyuanzhang16@stu.pku.edu.cn}
    \author{Daliang Xu}
    \authornote{Corresponding author.}
    \affiliation{\institution{Beijing University of Posts and Telecommunications}\city{Beijing}\country{China}}
    \email{xudaliang@bupt.edu.cn}
    \author{Yun Ma}
    \affiliation{\institution{Peking University}\city{Beijing}\country{China}}
    \email{mayun@pku.edu.cn}
    \author{Qing Li}
    \affiliation{\institution{Beijing University of Posts and Telecommunications}\city{Beijing}\country{China}}
    \email{qingli@bupt.edu.cn}
    \author{Mengwei Xu}
    \affiliation{\institution{Beijing University of Posts and Telecommunications}\city{Beijing}\country{China}}
    \email{mwx@bupt.edu.cn}
    \author{Shangguang Wang}
    \affiliation{\institution{Beijing University of Posts and Telecommunications}\city{Beijing}\country{China}}
    \email{sgwang@bupt.edu.cn}
    \author{Gang Huang}
    \affiliation{\institution{Peking University}\city{Beijing}\country{China}}
    \email{hg@pku.edu.cn}
    \author{Xuanzhe Liu}
    \affiliation{\institution{Peking University}\city{Beijing}\country{China}}
    \email{liuxuanzhe@pku.edu.cn}
    \renewcommand{\shortauthors}{Huang et al.}
    \input{sections/abstract}
    \maketitle
    \input{sections/introduction}
    \input{sections/background}
    \input{sections/overview}

    \input{sections/efficient_checkpointing}
    \input{sections/dynamic_recovery}
    \input{sections/implementation}
    \input{sections/evaluation}
    \input{sections/discussion}

    \input{sections/related_work}
    \input{sections/conclusion}

    \clearpage
    \bibliographystyle{ACM-Reference-Format}
    \bibliography{refs.bib}
\end{document}

%% file: sections/abstract.tex
\begin{abstract}
Executing vision-language models (VLMs) aboard low-Earth orbit satellites enables flexible processing of observation data without first downlinking it to the ground. On commercial off-the-shelf payloads, however, a request may span multiple execution windows: power and thermal constraints, fault-triggered reboots, and competing payload demands can interrupt inference before it completes. Each shutdown discards volatile activations and KV caches, forcing an unprotected request to repeat work after restart.

We present \sysname, a framework that preserves VLM inference progress across onboard interruptions through coordinated checkpointing and recovery. Its phase-aware checkpointing selects recovery boundaries according to the state lifetimes and computational characteristics of vision encoding, language-model prefill, and decoding, balancing storage traffic against recomputation. Its dependency-aware dynamic recovery schedules restoration from partially completed states across the layer and token-chunk dimensions. By overlapping checkpoint and weight reads with recomputation, it reduces restart latency; the recovery planner computes an optimal execution order under its scheduling model within 100\,ms.

Experiments on satellite payload hardware using on-orbit telemetry traces show up to a 4.07$\times$ end-to-end speedup and a 3.56$\times$ reduction in SSD write volume over the evaluated baselines. A 72-hour deployment on two low-Earth orbit satellites demonstrates that \sysname retains inference progress and resumes execution across repeated bus-enforced shutdowns.
\end{abstract}

%% file: sections/introduction.tex
\section{Introduction}
\label{sec:introduction}

Low-Earth orbit (LEO) satellites are emerging as powerful platforms for \textit{satellite computing}~\cite{asplos2020oec}, a new paradigm that shifts heavy data processing directly to orbit to overcome the intolerable latency of the traditional \textit{bent-pipe} architecture. 
Backed by the integration of commercial off-the-shelf (COTS) edge hardware (e.g., NVIDIA GPUs on Starcloud-1~\cite{starcloud2026starcloud1}), this high-performance orbital infrastructure is now physically accessible and cost-effective~\cite{mobicom2024cots, fcc2026spacex_orbital_data_centers, google2025suncatcher}. 
Leveraging these advanced platforms, Vision-Language Models (VLMs) are rapidly becoming the premier workload for dynamic space missions. 
Instead of relying on fragmented, task-specific models, VLMs offer unified, on-the-fly functional elasticity for effortless mission redirection~\cite{grace,geochat,wang2025viso}. 
Pioneering space-agency initiatives, such as NASA's Prithvi geospatial foundation model and the European Space Agency's (ESA) Phi-Sat-2 mission, have recently validated the feasibility of executing such foundation models directly in orbit~\cite{du2025firstonorbitdemo}, marking the dawn of intelligent satellite computing.

\para{Mismatch: sustained VLM execution vs. bus-enforced interruptions.}
On-orbit VLM inference requires sustained computation, while spacecraft-level control can interrupt payload execution before a request completes.

\textit{Long-running on-orbit VLM workloads.}
VLM inference on satellites can involve high-resolution images or multiple images within a single request.
These inputs can produce long visual-token sequences, increasing the computation needed to complete the request.
For example, Earth observation (EO) workloads can involve processing image sequences of up to 30 frames~\cite{ding2026longearth} and images reaching 44.93 megapixels~\cite{luo2025lrsvqa}.
Our in-orbit evaluation of Qwen3-VL 2B on \satelliteone and \satellitetwo shows that a typical 3-minute mission processes only 8 frames ($> 32K$ tokens).

\textit{Short and unpredictable inference survival windows.}
Onboard computing payloads face unpredictable interruptions that severely truncate execution windows. These abrupt halts primarily stem from three sources: 
(i) \textit{Bus-enforced hard shutdowns} due to dynamic power constraints or thermal violations~\cite{nasa_cii_hosted_payload_guidelines_2011}. For example, VLM inference on \satelliteone and \satellitetwo \footnote{We refer to the two satellites as \satelliteone and \satellitetwo throughout this paper.}  spikes the payload temperature from $45^\circ\text{C}$ to $55^\circ\text{C}$ in just 48\,s, leading to sudden thermal cutoffs that vary drastically depending on the satellite's orbital state.
(ii) \textit{On-orbit anomalies} (e.g., radiation-induced SEUs or software faults) that trigger the Fault Detection, Isolation, and Recovery (FDIR) system. FDIR forces sudden system reboots to clear errors; our private LEO satellite telemetry data reveals that in worst-case scenarios, such reboots occur as frequently as once per minute~\cite{evans2016opssatfdir}. 
(iii) \textit{Resource contention} from intermittent, high-priority payloads (e.g., bursty transceivers or laser transmitters) that preempt shared bus resources~\cite{fong2011formosat}.

To tackle this mismatch, we propose \sysname, an efficient and reliable VLM inference engine designed for satellite computing.
\sysname achieves transparent, minimal-overhead preservation of VLM execution progress by dynamically checkpointing essential states to onboard non-volatile storage (e.g., eMMC/SSD).
Upon rebooting, it swiftly reconstructs the context to resume inference.
Achieving this goal requires addressing two challenges:

\noindent $\bullet$ \textit{Checkpointing phase: fast checkpointing vs. efficient recovery.}
Preserving execution progress creates a dilemma: minimizing checkpointing I/O for active inference speed versus maximizing progress preservation for rapid recovery.
Fine-grained checkpointing (e.g., per-token/layer) captures complete progress but incurs substantial I/O overheads, whereas coarse-grained checkpointing (e.g., per-chunk) minimizes I/O but forces expensive recomputation of discarded intra-chunk states.
This trade-off manifests across all VLM phases.
For instance, in memory-bound decoding, saving per-token KV caches and hidden states ensures optimal recovery at prohibitive costs (>100\%), while merely saving token IDs eliminates I/O but forces full KV cache recomputation from scratch, introducing a latency of $>50$ seconds.

\noindent $\bullet$ \textit{Recovery phase: unavoidable weight-checkpoint I/O in unstructured resumption.}
Unlike traditional LLM context switching with uniform states (e.g., full prompt KV caches), arbitrary orbital bus-enforced interruptions leave an irregular, ``staircase-like'' inference progress state.
Resuming from this unstructured state complicates masking the coupled I/O overhead of reloading static weights alongside dynamic KV caches and hidden states from disparate positions.
Specifically, loading a layer's weights unlocks computation across multiple chunks, whereas loading its KV cache only enables that specific chunk.
With numerous valid starting points along this irregular boundary, finding the optimal execution sequence to maximize compute-I/O overlap under complex dependency constraints becomes a highly non-trivial scheduling challenge.

To solve these challenges, \sysname incorporates two key techniques:

\parabullet{Phase-aware checkpointing.}
To resolve the checkpointing dilemma, \sysname leverages the heterogeneous lifespans of VLM states (persistent KV caches vs. transient hidden states) and phase-specific bottlenecks to establish adaptive recovery boundaries: 
(i) \textit{Vision encoding (layer boundary):}
Because bidirectional self-attention couples all image patches, patch chunks cannot be finalized independently.
\sysname therefore persists the full-patch hidden state after each vision layer, using layer boundaries as recovery points.
(ii) \textit{LLM prefill (2D block boundary):}
Because prefill is compute-bound, checkpointing both KV Cache and hidden states minimally impacts active inference.
To mitigate SSD wear-out, \sysname groups operations into 2D blocks (execution grids spanning multiple sequence chunks and model layers), checkpointing transient hidden states exclusively at block boundaries. 
(iii) \textit{LLM decoding (multi-token boundary):}
Decoding is memory-bound, making per-token checkpointing prohibitively expensive.
Recognizing that the decoding recovery bottleneck is massive weight-loading I/O rather than lightweight recomputation, \sysname aggregates KV cache updates across multi-token boundaries.
This amortizes flash write overhead, as the slight recomputation increase is completely masked by the inevitable weight-loading I/O.

\parabullet{Dependency-aware dynamic recovery.}
To mitigate coupled I/O bottlenecks during unstructured resumption, we observe that despite recovery computation heavily outweighing I/O latency, strict dependencies still cause severe I/O stalls.
To maximize compute-I/O overlap, \sysname unlocks dependencies multi-dimensionally: 
(i) \textit{2D block-first recovery:}
Rigid layer- or chunk-first strategies only resolve single-dimension dependencies.
Instead, executing a 2D block immediately unlocks both horizontal (chunk) and vertical (layer) dependencies, triggering a computation cascade to mask concurrent SSD reads. 
(ii) \textit{Optimal scheduling plan:}
Selecting the optimal ready block to minimize recovery latency is critical, as each choice unlocks distinct downstream computation cascades to maximize I/O masking; yet this scheduling is NP-hard.
We dramatically prune this search space via two domain-specific properties: (a) \textit{deferring hidden state loads} to retain DRAM activations and avoid redundant SSD reads, and (b) \textit{packing layer-wise I/O} to coalesce weight reads into coarse-grained streams.
This enables \sysname to mathematically solve the optimal execution order within 100\,ms.

We implemented \sysname atop \texttt{transformers}~\cite{wolf2020transformerslib} and evaluated it on hardware testbeds mirroring two real satellite computing payloads: the NVIDIA Jetson Orin NX (16\,GB) and AGX Orin (64\,GB).
These represent the onboard platforms of \satelliteone and \satellitetwo.
Our evaluations are driven by five interruption traces: four derived from telemetry collected on \satelliteone and \satellitetwo, and one from the Dellingr satellite~\cite{hatfield2018dellingr}.
Using these scenarios, we run advanced edge-viable VLMs, including Qwen3-VL 2B/4B/8B and MiniCPM-V 4.5, on remote sensing QA datasets. 
Compared to three state-of-the-art baselines (two variants of a representative intermittent DNN execution runtime and a conventional KV cache context-saving approach), \sysname achieves up to a 4.07$\times$ end-to-end speedup.
Furthermore, it reduces checkpointing and recovery latency by up to 61.9\% and 85.6\%, respectively, while slashing SSD write volume by up to 3.56$\times$.
\textbf{To the best of our knowledge, we are the first academic work to successfully deploy and sustain VLM inference directly in orbit. We deployed \sysname on \satelliteone and \satellitetwo for a 72-hour in-situ earth observation case study. The deployment demonstrates that \sysname preserves progress and sustains VLM inference across repeated bus-enforced shutdowns.}

Our contributions are summarized as follows:

\begin{itemize}[leftmargin=1.25em,labelsep=0.45em,nosep]
\item  We are the first to identify the bottleneck of bus-enforced shutdowns in on-satellite VLM
inference.

\item We propose \sysname, the first efficient and reliable VLM inference system for satellite computing by introducing efficient checkpointing and recovery.

\item  We implement \sysname and deploy it on \satelliteone and \satellitetwo. Our evaluation shows that \sysname achieves up to a 4.07$\times$ end-to-end speedup, while a 72-hour in-orbit deployment demonstrates sustained VLM inference across repeated bus-enforced shutdowns.

\end{itemize}

%% file: sections/background.tex
\section{Background}
\label{sec:background}

\begin{figure}
    \centering
    \includegraphics{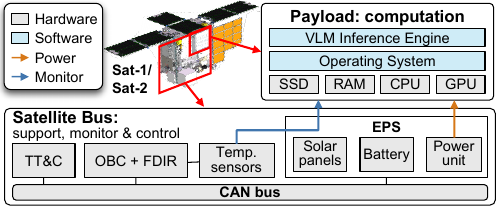}
    \Description{Satellite computing architecture with payload sensing and computing components connected to bus support and control subsystems.}
    \caption{Satellite architecture diagram. TT\&C $=$ Telemetry, Tracking, and Command; OBC $=$ On-Board Computer; FDIR $=$ Fault Detection, Isolation, and Recovery; EPS $=$ Electrical Power System. }
    \label{fig:satellite-diagram}
\end{figure}

\begin{figure*}[t]
    \centering
    \includegraphics[width=\textwidth]{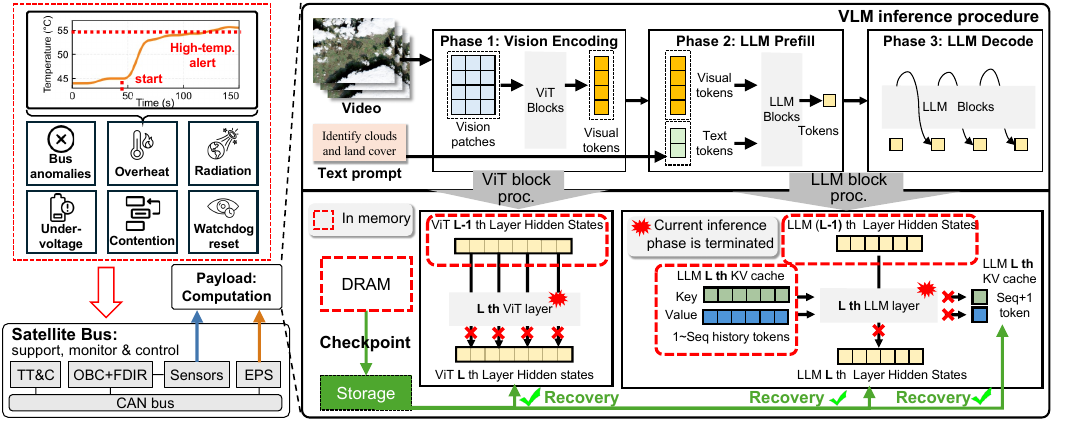}
    \Description{Three-panel workflow showing the payload enclosure temperature trace, thermal-triggered payload power off, spacecraft bus recovery control, and VLM inference stages from image tokens to KV cache.}
    \caption{The temperature trace measures payload enclosure temperature. TT\&C $=$ Telemetry, Tracking, and Command; OBC $=$ On-Board Computer; FDIR $=$ Fault Detection, Isolation, and Recovery; EPS $=$ Electrical Power System.}
    \label{fig:challenge}
\end{figure*}

\subsection{On-Satellite Vision-Language Models}
\label{sec:background_vlm}

\para{Satellite computing} shifts data processing and services directly to orbit.
This approach overcomes the severe communication bottlenecks of the traditional \textit{bent-pipe} architecture, where limited downlink bandwidth and brief contact windows make downlinking raw data for ground-side processing intolerably slow for modern space missions~\cite{asplos2020oec}. 
This orbital computing capability is enabled by moving beyond traditional radiation-hardened controllers and integrating commercial off-the-shelf (COTS) computing platforms—such as CPUs, GPUs, and FPGAs—into modern satellites~\cite{mobicom2024cots,asplos2023kodan,giuffrida2022phi1,rs2021fpgacnn}. 
Consequently, satellites are evolving from simple data collectors into orbital edge nodes capable of executing diverse tasks. 
Early examples include onboard filtering to discard cloudy images (Phi-Sat-1~\cite{giuffrida2022phi1}) or select high-value samples (Kodan~\cite{asplos2023kodan}). 
More recently, platforms like Starcloud-1 are pushing hardware boundaries by carrying data-center-class NVIDIA H100 GPUs in orbit, demonstrating that high-performance AI accelerators can now be deployed as standard satellite payloads.

As shown in \Fref{fig:satellite-diagram}, a typical satellite computing platform consists of a computing payload and a spacecraft bus that supports, monitors, and controls the payload.
The payload runs a VLM inference engine on top of an operating system, using SSD for persistent storage, RAM for volatile memory, and CPUs and GPUs for computation.
During inference, the engine loads model weights and input data from SSD into RAM and dispatches tensor operations to the processors, while intermediate tensors and runtime states remain in volatile memory.
Meanwhile, the spacecraft bus supplies payload power through the electrical power system (EPS), monitors payload temperature through sensors, and manages communication and fault protection through TT\&C and the onboard computer (OBC) with FDIR.
These bus subsystems communicate over a CAN bus and can disconnect payload power or trigger a reboot when power, thermal, or safety constraints require intervention, abruptly interrupting the ongoing inference.

\para{Vision-Language Models for LEO satellites.}
With advanced hardware in orbit, satellite computing tasks are rapidly shifting toward VLMs. 
For critical applications like Earth observation (EO), traditional fragmented, task-specific models are being replaced by VLMs, which offer a ``one-for-all'', prompt-driven framework that provides functional elasticity for dynamic missions without retraining overhead~\cite{du2025firstonorbitdemo}. 
Furthermore, driven by the massive expansion of LEO constellations, initiatives like Starcloud~\cite{feilden2024space} are actively constructing ``space data centers'' to deliver real-time, space-native VLM inference and token generation services directly to users.
Robust, sustained VLM execution is therefore essential for next-generation LEO systems.

\para{On-satellite VLM inference pipeline.}
A VLM is a multimodal model that converts images and text instructions into text outputs by combining visual encoding with language-model generation.
As illustrated in \Fref{fig:challenge}, after the on-board camera captures an image, the acquired visual data is fed into the VLM for inference, initiating a three-phase pipeline.
In the first phase (vision encoding), a Vision Transformer partitions the input image into patches and encodes them into visual tokens~\cite{iclr2021vit}.
During the second phase (LLM prefill), the model processes these visual tokens alongside the prompt tokens, establishing the attention state and initializing the Key-Value (KV) cache for the multimodal context.
Finally, in the LLM decoding phase, the language model generates the response autoregressively, producing one token at a time while continuously utilizing and updating the KV cache established during the prefill phase.
These phases involve different intermediate states, including visual tokens, prefill states, KV cache, and partial outputs.

\subsection{Bus-enforced Shutdowns in On-Satellite VLM Inference}
\label{sec:background_challenges}

Onboard computing payloads operate under strict spacecraft-wide power and safety constraints.
The spacecraft bus can abruptly disconnect payload power to preserve essential functions or protect hardware, regardless of the inference software's status~\cite{nasa_basics_power_distribution,evans2016opssatfdir}.
We term these platform-initiated interruptions \emph{bus-enforced shutdowns}, which can terminate VLM requests unpredictably. They primarily stem from three sources:

\parabullet{Dynamic power and thermal constraints.}
The electrical power system (EPS) autonomously sheds non-essential loads (e.g., computing payloads) when demand exceeds supply~\cite{nasa_basics_power_distribution, ali2026autonomousEPS}. 
This power availability fluctuates with solar illumination and battery aging, which alters discharge curves over time and makes supply highly non-linear. 
Furthermore, thermal limits force sudden shutdowns. As shown in \Fref{fig:challenge}, VLM inference on \satelliteone spikes the payload enclosure temperature from $45^\circ\text{C}$ to the $55^\circ\text{C}$ threshold in just 48\,s, triggering an immediate power-off. Driven by external orbital states, these shutdowns are entirely decoupled from VLM execution boundaries (e.g., prefill and decoding steps).

\parabullet{Hardware/software anomalies and FDIR interventions.}
The harsh space environment introduces stochastic anomalies, such as radiation-induced Single-Event Upsets (SEUs) or software faults (e.g., erroneously activating Sun-pointing during an eclipse). 
The onboard Fault Detection, Isolation, and Recovery (FDIR) system continuously monitors payload health. Upon detecting these anomalies or hardware violations (e.g., overcurrent)~\cite{evans2016opssatfdir}, FDIR forces sudden system reboots to clear errors. These fault-protection actions occur randomly, completely ignoring any payload execution schedule. Our private LEO satellite telemetry data reveals worst-case scenarios where these unpredictable reboots occur once per minute.

\parabullet{Resource contention from intermittent payloads.}
The computing payload shares the bus with high-priority, intermittent payloads (e.g., bursty transceivers)~\cite{fong2011formosat}. Their unpredictable activation---driven by ground station visibility or mission commands---aggressively consumes limited power and thermal budgets, often preempting the computing payload~\cite{fong2011formosat,ali2026autonomousEPS}. 
Together, these stochastic factors render payload shutdown timing and duration highly erratic and not practical to profile offline.

\para{On-satellite VLM inference needs fast checkpointing and recovery.}
As illustrated in \Fref{fig:challenge}, bus-enforced shutdowns can terminate any inference phase and discard its in-memory states, including ViT/LLM hidden states and the KV cache.
Restoring power alone therefore cannot resume execution; recovery requires checkpoints persisted in non-volatile storage.
Consequently, VLM inference requires low-overhead checkpointing and rapid recovery without advance knowledge of shutdown timing. 
Our scope addresses execution interruption and volatile state loss, assuming persistent weights and checkpoints in non-volatile storage remain intact (leaving radiation-induced persistent storage corruption to orthogonal hardware research).

%% file: sections/overview.tex
\section{Design Overview}
\label{sec:overview}

\begin{figure}
    \centering
    \includegraphics{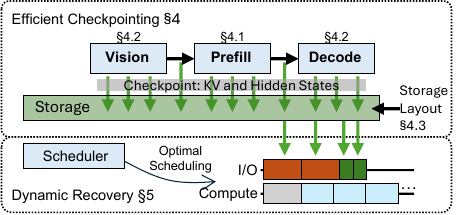}
    \Description{Satellite computing architecture with payload sensing and computing components connected to bus support and control subsystems.}
    \caption{Overview}
    \label{fig:overview}
\end{figure}

We present \sysname, an efficient and reliable VLM inference system designed for COTS satellites.
\Fref{fig:overview} shows the overview of \sysname.
To address the progress-loss problem caused by bus-enforced shutdowns, OrbitWeaver adopts checkpointing and recovery to make VLM inference resumable across power-off and reset events.
During inference, OrbitWeaver persists selected boundary states from volatile memory to SSD; after a shutdown, it validates and reloads the latest checkpoint to reconstruct the execution context and continue the unfinished request instead of restarting from scratch.

However, making checkpointing and recovery efficient on COTS satellites is nontrivial.
Checkpointing must preserve enough state to resume each VLM phase, but excessive persistence increases SSD traffic, energy use, and normal-path latency. 
Recovery must reload model weights and checkpointed states from SSD, but a poor recovery order exposes I/O stalls and leaves little useful computation time within each short execution window.
To address these two challenges, \sysname introduces two techniques.

\para{Efficient checkpointing.}
\sysname defines phase-aware recoverable units that preserve only the states needed for resumption while bounding runtime checkpoint overhead.
For LLM prefill, \sysname uses block-wise checkpointing over token chunks and layer groups, avoiding the hidden-state write cost of layer-wise checkpointing and the recomputation cost of chunk-wise checkpointing.
For vision encoding, \sysname checkpoints layer-boundary hidden states because bidirectional attention makes patch chunks invalid recovery units.
For decoding, \sysname batches KV cache checkpoints across generated tokens to avoid excessive small writes.
\sysname further uses phase-specific storage layouts to reduce I/O interference, avoid unnecessary recovery reads, and ensure crash-consistent checkpoint persistence.

\para{Dynamic recovery.}
\sysname reduces recovery latency by scheduling SSD reads and block computation over the checkpointed frontier.
Instead of following a fixed chunk-first or layer-first order, \sysname restores the states that unlock useful computation and overlaps later SSD reads with dependency-ready blocks.
By exploiting the regular dependency structure of VLM prefill, \sysname reduces recovery planning to layer-level I/O ordering and solves it with a lightweight dynamic programming algorithm.

%% file: sections/efficient_checkpointing.tex
\section{Efficient Checkpointing}
\label{sec:efficient_checkpointing}

\sysname preserves VLM progress across satellite shutdowns by persisting intermediate states at checkpoint boundaries.
Each boundary must expose enough state for correct resumption while limiting SSD traffic, energy cost, and runtime overhead. We identify three primary issues:

\begin{itemize}[leftmargin=*, itemsep=0pt, topsep=2pt, parsep=0pt]
    \item VLM phases expose different state dependencies, so a uniform checkpoint boundary cannot efficiently support vision encoding, prefill, and decoding;
    \item Checkpoint granularity creates a recomputation--write trade-off: persisting more states reduces replay after shutdowns, but increases checkpoint writes during normal execution.
    \item Checkpoint persistence must be storage-aware: fragmented writes increase runtime overhead and SSD wear, while storing hidden states and KV tensors in an interleaved layout forces recovery to read unnecessary data.
\end{itemize}

To address these issues, OrbitWeaver designs phase-aware recoverable units.
It uses block-wise checkpointing for prefill (\Sref{sec:block_prefilling}), phase-specific checkpointing for vision and decoding (\Sref{sec:checkpointing_vision_decoding}), and an SSD layout that supports efficient and crash-consistent checkpoint persistence (\Sref{sec:storage_layout}).

\subsection{Block-wise Prefill Checkpointing}
\label{sec:block_prefilling}

When prefill completes without a shutdown, preserving its KV cache is sufficient for the later decoding phase.
However, a satellite reset may occur in the middle of prefill, discarding all in-memory progress before the final KV cache is produced.
Therefore, OrbitWeaver must checkpoint prefill at finer granularity, along its layer and token dimensions. This leads to two straightforward checkpoint placements.

\begin{figure}[t]
    \centering
    \includegraphics{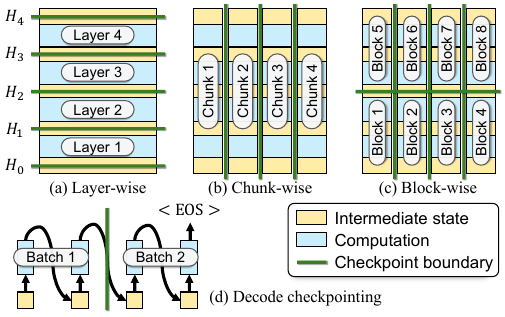}
    \Description{Comparison of layer-wise, chunk-wise, and block-wise prefill checkpointing, with computation blocks, intermediate states, and checkpoint boundaries.}
    \caption{\sysname 's  checkpointing strategies.}
    \label{fig:checkpointing_diagram}
\end{figure}

\begin{figure}[t]
    \centering
    \includegraphics{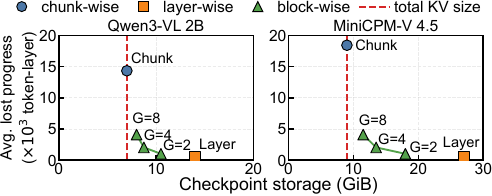}
    \Description{Trade-off between checkpointing overhead and lost computation for different block sizes.}
    \caption{Storage--recomputation trade-off of different checkpointing granularities.}    
    \label{fig:checkpoint_tradeoff}
\end{figure}

\para{Na\"ive solution: Layer-wise checkpointing.}
A natural design is to follow the LLM's layer-by-layer execution order and place a checkpoint boundary after each layer, as shown in \Fref{fig:checkpointing_diagram}(a).
However, this layer-level granularity comes at a high checkpointing cost.
Since recovery from a layer boundary requires the input to the next layer, \sysname must persist the full-sequence hidden state at every layer boundary, whose size scales with the multimodal input length, including both visual tokens and prompt tokens.
As shown in \Fref{fig:checkpoint_tradeoff}, layer-wise checkpointing minimizes lost progress but requires the largest checkpoint storage: about $14.34$ GiB for Qwen3-VL 2B and $27.65$ GiB for MiniCPM-V 4.5.

\para{Na\"ive solution: Chunk-wise checkpointing.}
Another design is to partition the input sequence into fixed-size chunks and place a checkpoint boundary after each chunk completes all $L$ layers, as shown in \Fref{fig:checkpointing_diagram}(b).
This reduces checkpoint traffic: after a chunk finishes all layers, the checkpoint only needs to preserve the KV tensors produced by the chunk, rather than full-sequence hidden states after every layer.
As shown in \Fref{fig:checkpoint_tradeoff}, chunk-wise checkpointing keeps storage near the total KV size, but incurs the highest recomputation overhead: about $14.34 \times 10^3$ and $18.43 \times 10^3$ token-layers for Qwen3-VL 2B and MiniCPM-V 4.5, respectively.

The comparison above reveals a structural property of LLM prefill: KV tensors are long-lived context; once produced, they are reused by later chunks and by decoding. Hidden states are short-lived boundary states; they are only needed to feed the next layer group after recovery.
Layer-wise checkpointing over-preserves hidden states, while chunk-wise checkpointing under-preserves them and cannot resume inside a chunk.
\sysname exploits this separation by maintaining a minimal recovery frontier: it accumulates KV tensors for completed computation and materializes hidden states only at sparse layer-group boundaries.

\para{\sysname design: Block-wise checkpointing.}
Based on this frontier view, \sysname partitions prefill along both the token and layer dimensions.
It splits the input sequence into chunks of $C$ tokens and groups the $L$ LLM layers into groups of $G$ consecutive layers, forming a two-dimensional grid of recoverable units (\Fref{fig:checkpointing_diagram}(c)).
Each unit computes $G$ layers over $C$ tokens.
When a unit completes, \sysname persists the KV tensors produced by the unit and the hidden states at the output boundary of the layer group. 

This design preserves only the states needed at the recovery frontier.
Compared with layer-wise checkpointing, it avoids persisting full-sequence hidden states after every layer.
Compared with chunk-wise checkpointing, it can resume inside a chunk and replay at most the unfinished $G \times C$ unit instead of the entire chunk across all $L$ layers.
The choice of $G$ and $C$ determines where \sysname operates on the storage--recomputation trade-off.
Smaller units reduce recomputation after a reset but increase transfer, write, and metadata overhead; larger units reduce checkpointing overhead but lose more computation after a shutdown.
\sysname selects $G$ and $C$ through lightweight offline calibration on the LRS-VQA~\cite{luo2025lrsvqa} for each model-device pair.
The calibration profiles a small set of candidate configurations on representative prefill inputs and selects the one that satisfies a user-defined checkpoint overhead budget while bounding the replay cost after a reset.
With $G=4$, \sysname reduces lost progress to about $2.04 \times 10^3$ token-layers for both models, with checkpoint sizes of $8.96$ GiB for Qwen3-VL 2B and $13.82$ GiB for MiniCPM-V 4.5 (\Fref{fig:checkpoint_tradeoff}).

\subsection{Checkpointing Vision and Decode}
\label{sec:checkpointing_vision_decoding}

Beyond LLM prefill, the other VLM phases expose different recovery boundaries.
Vision encoding is dominated by layer-by-layer bidirectional attention, while decoding advances autoregressively one token at a time.
\sysname therefore uses phase-specific checkpointing policies for these two phases.

\para{Layer-wise vision checkpointing.}
Unlike LLM prefill, whose causal attention allows completed token chunks to be finalized independently, vision encoding uses bidirectional self-attention over all patches~\cite{iclr2021vit}.
Since each patch output depends on the full patch sequence, patch chunks cannot serve as valid recovery units.
Once a vision layer completes, its full-patch hidden state becomes the complete input to the next layer, so \sysname checkpoints vision encoding at layer boundaries by persisting this state.

\para{Batched decode checkpointing.}
Decoding advances autoregressively, generating one output token at a time.
Each generated token appends new entries to the KV cache, and future tokens depend on these entries.
Therefore, decoding only checkpoints the KV cache as its recoverable progress state, while hidden states are transient within each decode step and need not be persisted.
Checkpointing after every token minimizes replay but is inefficient: each token adds only a small KV slice, so per-token checkpoints cause frequent device-to-host transfers and fragmented SSD writes. 
This is especially costly for memory-bound decoding~\cite{zhong2024distserve}, where extra checkpoint writes compete with incremental KV cache access.
\sysname therefore batches decode checkpoints.
It accumulates the KV slices for $D$ generated tokens and persists them as one checkpoint unit, together with the generated token IDs and the current decode position.
If a reset occurs between two decode checkpoints, recovery reloads the latest persisted KV cache and replays at most $D$ decode steps.

\subsection{I/O-Efficient Storage Layout}
\label{sec:storage_layout}

The checkpointing policies above define logical recoverable units; \sysname
then lays out their persisted states according to each phase's access
pattern to reduce I/O overhead and ensure crash consistency.

\begin{figure}[t]
    \centering
    \includegraphics{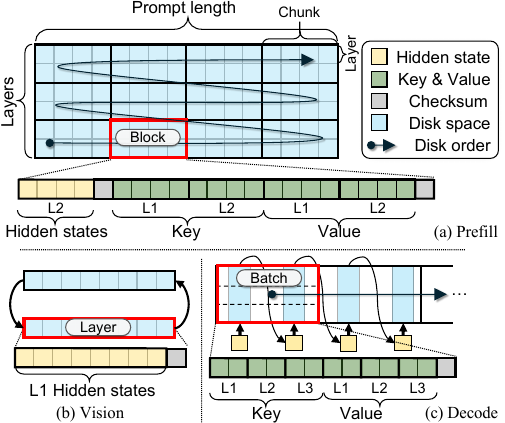}
    \Description{Storage layout for persisted checkpoints across prefill, vision, and decoding, including KV cache and hidden-state buffers.}
    \caption{Checkpoint storage layout for prefill, vision, and decode.}
    \label{fig:storage_layout}
\end{figure}

\para{Prefill layout.}
This layout is optimized for the asymmetric access patterns of checkpointing and recovery. 
For block-wise prefill checkpoints, \sysname stores the states produced at each completed block boundary as one contiguous record (\Fref{fig:storage_layout}(a)), which turns each checkpoint into a sequential write.
Each record separates boundary hidden states from the block's multi-layer KV tensors: the former resumes the next layer group, while the latter serves later chunks and decoding.
This separation lets recovery restore only the needed hidden-state or KV region, avoiding unnecessary I/O.
\sysname further stores KV tensors in layer-major order, allowing a layer's prefix KV to be restored through sequential reads.

\para{Vision layout.}
For vision checkpoints, \sysname reduces storage overhead by persisting only the layer-boundary hidden states (\Fref{fig:storage_layout}(b)).
Each hidden-state checkpoint is written as one contiguous record.
To tolerate resets during checkpoint writes, \sysname alternates between two aligned buffers: the new buffer is marked valid only after its write completes, while the previous buffer remains valid during the write.
If a reset interrupts the current write, recovery selects the newest complete buffer.

\para{Decode layout.}
Decode checkpoints must handle a dynamically growing output sequence (\Fref{fig:storage_layout}(c)).
Since future token length is unknown, \sysname uses an append-only layout instead of preallocating fixed regions for layers or positions.
Each decode checkpoint appends the KV slices generated by one decode batch as a single record.
Within the record, keys from all layers are written contiguously, followed by values from all layers, turning batched KV updates into sequential writes.
This layout grows naturally with generation while preserving efficient checkpoint persistence.

\para{Checksum and crash consistency.}
System resets can happen during checkpoint writes, leaving partially written checkpoint data on storage.
To handle this, \sysname appends a checksum-protected commit footer to each checkpoint record.
The footer contains the record length, checkpoint index, and payload checksum, and the record is considered committed only after this footer is durably written.

%% file: sections/dynamic_recovery.tex
\section{Dynamic Recovery}
\label{sec:dynamic_recovery}

After the satellite recovers from a shutdown, the VLM inference must restore the checkpointed state from the SSD and resume computation.
The latency of this process greatly affects the end-to-end completion time of a request.

\begin{figure}[t]
    \centering
    \includegraphics[width=\linewidth]{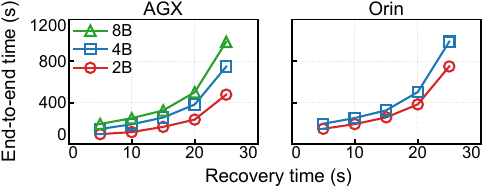}
    \Description{Two line plots showing that end-to-end time increases with recovery time on AGX and Orin. AGX reports Qwen3-VL 2B, 4B, and 8B, while Orin reports Qwen3-VL 2B and 4B.}
    \caption{Impact of recovery latency on interrupted VLM inference with a 64K-token request.}
    \label{fig:recovery_latency_impact}
\end{figure}

\para{Fast recovery is critical.}
To motivate the need for fast recovery, we run a controlled experiment with Qwen3-VL 2B/4B/8B on Jetson AGX Orin and Jetson Orin NX.
The workload is a 64K-token remote-sensing VQA request constructed from our evaluation datasets.
We fix the checkpointed execution progress and reset pattern, and vary the recovery time paid at the beginning of each execution window.
\Fref{fig:recovery_latency_impact} shows that increasing recovery time from 5\,s to 25\,s raises end-to-end completion time by up to 5.04$\times$ and 5.56$\times$ on AGX and Orin, respectively.
Therefore, \sysname minimizes exposed recovery I/O before computation resumes.

In this section, we first show why fixed recovery orders create avoidable I/O stalls (\Sref{sec:static_recovery}).
Then we formulate recovery as an I/O--computation scheduling problem (\Sref{sec:problem_formulation}).
Finally, we illustrate the algorithm to find the optimal schedule (\Sref{sec:optimal_recovery_plan}).

\subsection{Dependency-Aware Recovery}
\label{sec:static_recovery}

Following the execution order of existing LLM prefill implementations, the most direct recovery strategy is to replay the remaining prefill blocks in the same static order.
The order of existing implementations is typically chunk-first or layer-first. However, recovery optimizes a different objective from normal inference: it must load weights and checkpoints and start computation as early as possible.
To minimize recovery latency, we need to minimize \textit{exposed I/O}, i.e., the time computation stalls waiting for unfinished reads.

\begin{figure}[t]
    \centering
    \includegraphics{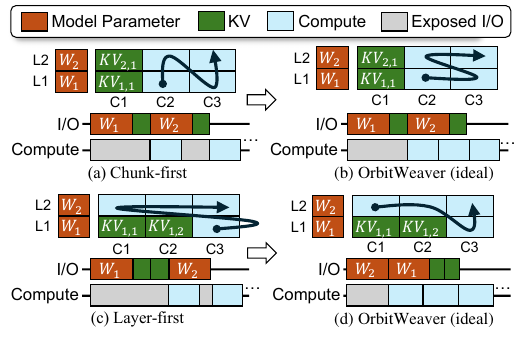}
    \Description{Four examples of static and optimal recovery orders over two layers and three chunks. Orange boxes denote model parameters, dark green boxes denote KV checkpoints, and arrows denote compute order.}
    \caption{Fixed recovery orders expose avoidable I/O stalls.}
    \label{fig:static_recovery}
\end{figure}

\parabullet{Chunk-first static recovery.}
In \Fref{fig:static_recovery}(a), the first chunk has completed before the shutdown.
A chunk-first order resumes from $(L_1,C_2)$ and proceeds to $(L_2,C_2)$.
This switch requires $W_2$ and $KV_{2,1}$; any part of these reads not hidden by the computation of $(L_1,C_2)$ becomes exposed I/O.
Thus, chunk-first recovery can stall when it switches to a higher layer before the required layer state has finished loading.

\parabullet{Layer-first static recovery.}
In \Fref{fig:static_recovery}(c), layer $L_1$ has completed through chunk $C_2$ before the shutdown.
A layer-first order starts with $(L_1,C_3)$, which requires $W_1$, $KV_{1,1}$, and $KV_{1,2}$ before any computation can run, exposing them at the beginning of recovery.
Thus, layer-first recovery can stall by choosing a block whose inputs are not yet available, even when other blocks are ready to execute.

\parabullet{Ideal dependency-aware recovery.}
Ideally, we can avoid these stalls by following data readiness rather than a fixed order. 
In \Fref{fig:static_recovery}(b), recovery keeps computing ready blocks in $L_1$ to hide the reads needed by $L_2$.
In \Fref{fig:static_recovery}(d), recovery first executes the ready prefix of $L_2$ to hide the reads needed later by $(L_1,C_3)$ ($(L_2,C_3)$ must wait for $(L_1,C_3)$).
The ideal recovery order therefore depends on which weights, KV cache, and hidden states are available, motivating a constrained I/O--computation scheduling problem.

\subsection{Recovery as a Scheduling Problem}
\label{sec:problem_formulation}

To realize dependency-aware recovery, \sysname formulates recovery as a scheduling problem over I/O and compute tasks to maximize I/O-compute overlap.

\begin{figure}[t]
    \centering
    \includegraphics{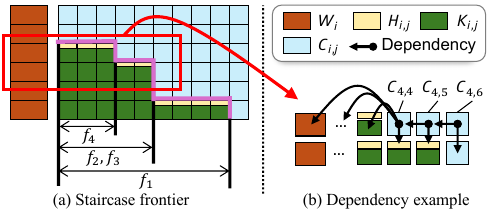}
    \Description{Frontier dependency diagram. Panel (a) shows a staircase frontier with markers for f1, f2, f3, and f4. Panel (b) shows dependency examples for three compute tasks in the fourth layer.}
    \caption{Frontier Dependency. (a) A concrete staircase frontier after a shutdown, with $f_1$, $f_2$, $f_3$, and $f_4$ marking the completed prefix of each layer. (b) Dependency examples for three compute tasks, $\mathsf{C}_{4,4}$, $\mathsf{C}_{4,5}$, and $\mathsf{C}_{4,6}$.}
    \label{fig:frontier_dependency}
\end{figure}

\para{Task definition.}
Let $N$ and $L$ be the total number of chunks and layers after layer grouping (\Sref{sec:block_prefilling}), respectively.
There are four types of tasks as summarized in \Tref{tab:recovery_tasks}.
LLM's causal attention requires each block to depend on: (i) earlier blocks in the same layer; (ii) the same block from the previous layer.
Thus, any computed block implies all blocks left and below it in the prefill grid are already computed.
Intuitively, the completed blocks form a staircase frontier in the prefill grid, as shown by the pink polyline in \Fref{fig:frontier_dependency}(a).
Formally, let $f_i$ denote the number of completed blocks in layer $i$, with $f_0=N$; then $N=f_0 \ge f_1 \ge \cdots \ge f_L \ge 0$.
Here, $f_1=9$, $f_2=f_3=5$, and $f_4=3$.
Blocks with $j \le f_i$ can be restored from disk, while those with $j > f_i$ must be computed.

\para{Task dependencies.}
Recovery scheduling must respect the data dependencies among I/O and compute tasks.
A compute task is ready only when three inputs are available: (i) the layer weights; (ii) the KV cache of preceding chunks in the same layer; and (iii) the hidden state from the previous layer.
These input requirements translate into the following task dependencies for each compute task $\mathsf{C}_{i,j}$ with $f_i<j\le N$:
\begin{equation*}
    \refstepcounter{equation}
    P(\mathsf{C}_{i,j}) = \mathsf{W}_i,\ 
    \left\{
        \begin{array}{@{}ll@{}}
        \mathsf{K}_{i,1},\dots,\mathsf{K}_{i,f_i},\mathsf{H}_{i-1,j},   & j=f_i+1\le f_{i-1},\\
        \mathsf{K}_{i,1},\dots,\mathsf{K}_{i,f_i},\mathsf{C}_{i-1,j},   & j=f_i+1>f_{i-1},\\
        \mathsf{C}_{i,j-1},\mathsf{H}_{i-1,j},                          & f_i+1 < j \le f_{i-1},\\
        \mathsf{C}_{i,j-1},\mathsf{C}_{i-1,j},                          & \text{otherwise}.\makebox[0pt][l]{\hspace{2.5em}\textup{(\theequation)}}
        \end{array}
    \right.
    \label{eq:compute_task_dependencies}
\end{equation*}

\Fref{fig:frontier_dependency}(b) illustrates this dependency structure: $\mathsf{C}_{4,4}$, $\mathsf{C}_{4,5}$, and $\mathsf{C}_{4,6}$ correspond to the first, third, and fourth cases in \Eref{eq:compute_task_dependencies}, respectively.

\begin{table}[t]
    \centering
    \small
    \caption{Tasks in the dynamic recovery scheduling model.}
    \begin{tabular}{@{}lccc@{}}
        \toprule
        Task & Symbol & Type & Execution time \\
        \midrule
        Load model weights & $\mathsf{W}_i$ & I/O & $\tau_{\mathsf{W}}$ \\
        Load hidden states & $\mathsf{H}_{i,j}$ & I/O & $\tau_{\mathsf{H}}$ \\
        Load KV cache & $\mathsf{K}_{i,j}$ & I/O & $\tau_{\mathsf{K}}$ \\
        Execute block & $\mathsf{C}_{i,j}$ & Compute & $\tau_{\mathsf{C}}$ \\
        \bottomrule
    \end{tabular}
    \label{tab:recovery_tasks}
\end{table}

\para{Problem formulation.}
With the dependencies between tasks, recovery becomes a constrained I/O--computation scheduling problem.
Let $\mathcal{I}$ be all the load tasks (i.e., $\mathsf{W}_i$, $\mathsf{H}_{i,j}$, and $\mathsf{K}_{i,j}$), $\mathcal{C}$ be all the compute tasks (i.e., $\mathsf{C}_{i,j}$), and $\mathcal{T}=\mathcal{I}\cup\mathcal{C}$ be the set of all tasks.
A recovery schedule is equivalent to assigning each task $x\in\mathcal{T}$ a start time $s(x)$.
Let $\tau_x$ be the fixed execution time of the task type of $x$ from \Tref{tab:recovery_tasks}; then task $x$ occupies the interval $J_x=\left(s(x),s(x)+\tau_x\right)$.
A valid schedule must satisfy three conditions: (i) each task starts after its predecessors finish; (ii) I/O tasks do not overlap; and (iii) compute tasks do not overlap:
\begin{equation}
    \left\{
    \begin{array}{ll}
        s(x) \ge s(u)+\tau_u,   & x\in\mathcal{T},\ u\in P(x),\\
        J_x\cap J_y=\emptyset,  & x\ne y,\ x,y\in R,\ R\in\{\mathcal{I},\mathcal{C}\}.
    \end{array}
    \right.
    \label{eq:valid_schedule_constraints}
\end{equation}

Let $(c_1,\ldots,c_m)$ be the compute tasks ordered by their start times.
The exposed I/O between $c_{k-1}$ and $c_k$ is $s(c_k)-(s(c_{k-1})+\tau_{c_{k-1}})$ ($k \geq 2$).
Together with the initial wait $s(c_1)$, we derive the exposed I/O of schedule $s$ and the goal:
\begin{equation}
    \begin{aligned}
        & E(s) = s(c_1) + \sum_{k=2}^{m} \bigl(s(c_k)-(s(c_{k-1})+\tau_{c_{k-1}})\bigr),\\
        & \text{Goal:}\quad \text{find} \quad s^\star = \arg\min_{s} E(s).
    \end{aligned}
    \label{eq:optimal_recovery_schedule}
\end{equation}

\subsection{Optimal Recovery Plan}
\label{sec:optimal_recovery_plan}

The optimization problem above is an instance of precedence-constrained two-machine scheduling.
The general case of this problem is NP-hard.
Our problem size is $|\mathcal{T}|=O(NL)$, where $N$ grows with the request input length, so an algorithm exponential in $|\mathcal{T}|$ is unacceptable.

Fortunately, VLM prefill has a regular dependency structure that allows us to develop a specialized algorithm.
We find three opportunities to reduce the computing complexity:

\parabullet{Deferring loading hidden states.}
First, each checkpointed hidden state $\mathsf{H}_{i,j}$ is used only by the block directly above it, $\mathsf{C}_{i+1,j}$.
Therefore, \sysname can defer loading $\mathsf{H}_{i,j}$ until $\mathsf{C}_{i+1,j}$ starts and pack this read into the compute task.

\parabullet{Greedily scheduling compute tasks.}
Second, once the order of I/O tasks is fixed, any work-conserving order of ready compute tasks yields the same exposed I/O, so the compute side can schedule them greedily.
With each hidden-state read packed into its corresponding compute task, each packed compute task hides $\tau_{\mathsf{C}}-\tau_{\mathsf{H}}$ of I/O.

\parabullet{Packing I/O of the same layer.}
Third, because no compute task in a layer can start before the layer weights and all preceding KV blocks are loaded, delaying these reads until immediately before the last one does not worsen the schedule.
By repeatedly applying this operation, the I/O tasks of the layer can be made consecutive, with arbitrary internal order.
So we pack $W_i$ and all $K_{i,j}$ in layer $i$ into a single layer I/O operation, whose I/O cost is denoted by $\rho_i$.

Given these properties, we reduce the original $O(NL)$ block-level recovery problem to choosing an order for $O(L)$ layer I/O operations.
Each loaded layer incurs I/O time but unlocks compute tasks that can mask later I/O.
We model this process by growing the loaded set from $S_0=\emptyset$: if layer $\pi_m$ is loaded at step $m$, then $S_m=S_{m-1}\cup\{\pi_m\}$ for $1\le m\le L$; let $R(S)$ be the total I/O time spent on layers in $S$ and $C(S)$ the total compute time they make available.
The total exposed I/O is
\begin{equation}
    \label{eq:exposed_io_layer_order}
    E(\pi) = \max_{1 \le m \le L} \left(R(S_{m})-C(S_{m-1})\right)_+,
\end{equation}
where $(x)_+=\max(x,0)$.

\para{Finding the optimal layer order.}
Eq.~\eqref{eq:exposed_io_layer_order} shows that once the current loaded set $S_m$ is fixed, all future loading costs and decisions depend only on $S_m$, regardless of the specific layer order used to reach it.
Therefore, we can use subset dynamic programming to find the optimal layer loading order.
Let $F[S]$ be the minimum exposed I/O achievable after exactly the layers in set $S$ have been loaded. We have:
\begin{equation}
    \label{eq:dp_recurrence}
    \left\{
    \begin{aligned}
        F[S]            &= \min_{\substack{i\in S\\ S'=S\setminus\{i\}}} \max\left\{F[S'],\, R(S)-C(S')\right\} && \forall S\neq\emptyset,\\
        F[\emptyset]    &=0.
    \end{aligned}
    \right.
\end{equation}

\sysname enumerates loaded sets $S$ by increasing cardinality, records the best predecessor for each transition, and backtracks to obtain the optimal layer order.
The algorithm uses $O(2^L)$ space and $O(L\cdot2^L)$ time for $L$ layers; for Qwen3-VL 8B, it finishes within 100 ms when the layer group size $G\ge2$.

%% file: sections/implementation.tex
\section{Implementation}
\label{sec:implementation}

We implement \sysname with 9,617 lines of Python and 1,741 lines of CUDA/C++ on top of \texttt{transformers}.
The prototype processes one request at a time on a single GPU, matching our satellite payload setting.
\sysname is not tied to \texttt{transformers}: its checkpointing and recovery mechanisms operate on VLM execution states and can be integrated into engines such as vLLM~\cite{kwon2023vllm} and SGLang~\cite{zheng2024sglang}.

\para{Context compression.}
During long decoding phases, repeated resets may force recovery to reload the full prefill KV context many times.
\sysname therefore optionally compresses the persisted KV cache after prefill and stores it in a packed layout for decode recovery, reducing decode-stage recovery latency by up to $5.02\times$.
Our prototype supports KIVI~\cite{liu2024kivi} and TurboQuant~\cite{zandieh2025turboquant}, and can run KV cache quantization asynchronously without blocking decoding.

\para{Checksumming and crash consistency.}
To tolerate resets during checkpoint writes, \sysname protects each checkpoint record with a checksum-protected commit footer.
The footer is written atomically using \texttt{pwritev2()} on an \texttt{O\_DIRECT} file, aligned to the NVMe AWUPF~\cite{nvme_nvm_command_set_1_0c} of the satellite SSD.
After \texttt{fdatasync()}, recovery accepts only records with a valid footer and matching checksum, discarding incomplete or corrupted records.

%% file: sections/evaluation.tex
\section{Evaluation}
\label{sec:evaluation}

\subsection{Experiment setup}
\label{sec:evaluation:setup}

\begin{figure*}[t]
    \centering
    \includegraphics{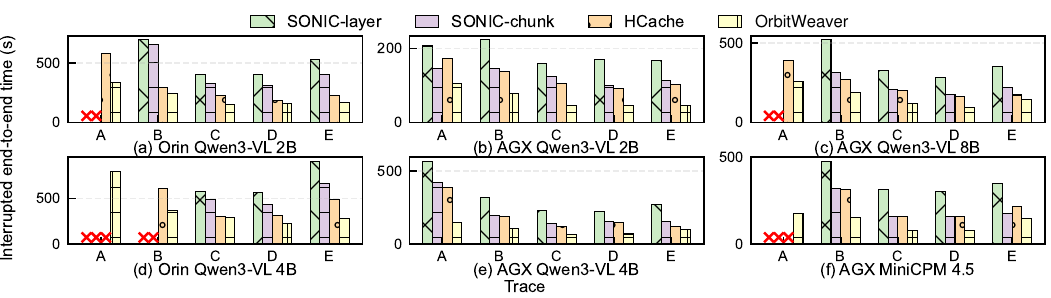}
    \Description{Six bar charts compare interrupted end-to-end time across checkpointing baselines and OrbitWeaver for different interruption traces, devices, and VLMs.}
    \caption{End-to-end performance under different setups}
    \label{fig:end_to_end_latency}
\end{figure*}

\para{Hardware setup.}
We test \sysname on two NVIDIA Jetsons with different compute capabilities: Jetson Orin NX (8-core ARM CPU, 1.9 TFLOPS Ampere GPU, 16\,GB memory) and Jetson AGX Orin (12-core ARM CPU, 5.3 TFLOPS Ampere GPU, 64\,GB memory), mirroring the computing payloads of \satelliteone and \satellitetwo. Both devices run Linux for Tegra R36.4.3.

\para{Models and dataset.}
We conduct experiments using four typical edge-sized VLMs: Qwen3-VL 2B/4B/8B and MiniCPM-V 4.5 (8B).
All models run W4A16 inference using GPTQ quantization and Marlin kernels.
We use LRS-VQA~\cite{luo2025lrsvqa} as a representative remote-sensing VQA workload; its images have an average resolution of $7{,}099 \times 6{,}329$ pixels.
To evaluate checkpointing and recovery performance, we construct VLM requests with specified lengths using image and text inputs from this dataset.

\para{Baselines.}
We compare \sysname with 4 baselines, including a naive no-checkpoint implementation, two variants of SONIC~\cite{asplos2019sonic} that targets general DNN inference on intermittent systems, and HCache that targets datacenter LLM context recovery.
Since SONIC does not support GPU and HCache does not open-source their code, we reimplement their techniques as described in their papers on \texttt{transformers} framework.
\begin{itemize}[leftmargin=*, itemsep=0pt, topsep=0pt]
    \item \textbf{No checkpoint}: naive VLM inference without checkpointing or recovery.
    \item \textbf{SONIC-layer}~\cite{asplos2019sonic}: layer-grained checkpointing that saves both KV cache and hidden states.
    \item \textbf{SONIC-chunk}~\cite{asplos2019sonic}: chunk-grained checkpointing that saves KV cache during chunk-first prefill.
    \item \textbf{HCache}~\cite{gao2025hcache}: hidden-state checkpointing with append-based storage and KV reconstruction during recovery.
\end{itemize}

\para{Metrics.} We mainly measure the end-to-end inference duration, recovery latency, inference latency, total SSD write amount, device level write amplification (DLWA), and energy consumption.
We obtain DLWA by collecting I/O traces at runtime and replaying them in MQSim~\cite{tavakkol2018mqsim}.
Energy is measured with \texttt{tegrastats}.
Before measuring recovery latency, we clear the OS page cache to evaluate cold-cache recovery performance.

\subsection{End-to-end performance}
\label{sec:evaluation:end_to_end}

\begin{figure*}[t]
    \centering
    \includegraphics{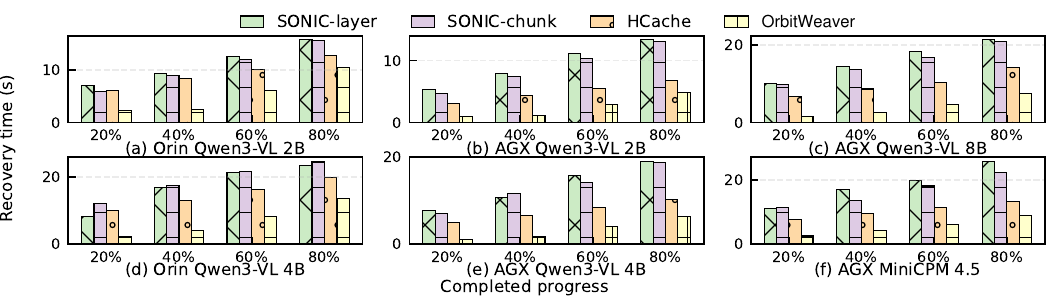}
    \Description{Six bar charts compare recovery time across checkpointing baselines and OrbitWeaver at different completed-progress points, devices, and VLMs.}
    \caption{Recovery latency under different setups.}
    \label{fig:recovery_latency}
\end{figure*}

We first compare the end-to-end performance of \sysname with SONIC-layer, SONIC-chunk, and HCache under Traces A--E (\Tref{tab:interruption_traces}).
We use interruption traces derived from telemetry collected on \satelliteone and \satellitetwo, together with a trace from Dellingr~\cite{hatfield2018dellingr}.
\Tref{tab:interruption_traces} summarizes their execution windows.

\begin{table}[t]
    \centering
    \small
    \caption{Interruption traces used in evaluation.}
    \label{tab:interruption_traces}
    \begin{tabular}{@{}lccccc@{}}
        \toprule
        Trace & A & B & C & D & E \\
        \midrule
        Satellite & \satelliteone & \satelliteone & \satellitetwo & \satellitetwo & Dellingr~\cite{hatfield2018dellingr} \\
        Mean window (s) & 36.4 & 42.3 & 121.7 & 108.2 & 63.0 \\
        \bottomrule
    \end{tabular}
\end{table}

On Orin, we evaluate Qwen3-VL 2B/4B. On AGX, we evaluate Qwen3-VL 2B/4B/8B and MiniCPM-V 4.5.
We use a 32K-token context on Orin and a 64K-token context on AGX (representative of the visual contexts for LRS-VQA images), with each request generating 64 tokens during decoding.
Part of each context is occupied by image tokens, and the rest is filled with text tokens.
For each model-device pair, we replay each trace and measure the time required to complete a VLM request across interruptions.
At each shutdown, we terminate the inference process and clear its volatile state; inference resumes from the latest checkpoint in the next execution window.
We repeat each trace until the request completes or reaches a limit of 20 full replays, after which it is marked as incomplete.
\Fref{fig:end_to_end_latency} shows the resulting completion times.
\sysname completes every evaluated request and achieves the lowest end-to-end completion time.
For completed baseline runs, \sysname achieves speedups of up to $4.07\times$, $2.89\times$, and $2.67\times$ over SONIC-layer, SONIC-chunk, and HCache, respectively.
The corresponding geometric-mean speedups are $2.95\times$, $1.98\times$, and $1.65\times$.

\sysname reduces both checkpointing and recovery overhead.
SONIC-layer writes KV cache and full-sequence hidden states at layer boundaries, while SONIC-chunk reduces these writes but may replay an unfinished chunk after a shutdown.
HCache checkpoints hidden states and later reconstructs the KV cache, overlapping reconstruction with state reads.
\sysname's phase-aware checkpointing (\Sref{sec:efficient_checkpointing}) uses block boundaries to avoid per-layer full-sequence hidden-state writes while bounding lost work within a chunk.
During recovery, the SONIC variants reload model weights and accumulated KV state before resuming prefill, while HCache restores the saved context through state loading and KV reconstruction.
\sysname's dynamic recovery (\Sref{sec:dynamic_recovery}) overlaps dependency-ready prefill blocks with other reads, reducing exposed I/O and reaching the next checkpoint sooner.

Crosses in \Fref{fig:end_to_end_latency} mark requests incomplete after 20 trace replays.
In the short windows of Traces A and B, recovery cannot finish in time to resume useful computation.
Each shutdown interrupts recovery, forcing the next window to restart from the same checkpoint without progress.
In Trace A, only \sysname completes Qwen3-VL 4B on Orin and MiniCPM-V 4.5 on AGX; in Trace B, the SONIC variants still fail on the former.
All baselines complete under the longer windows of Traces C--E, while \sysname remains faster by reducing checkpointing and recovery overhead.

\subsection{Recovery latency}
\label{sec:evaluation:recovery_latency}

\begin{figure*}[t]
    \centering
    \includegraphics{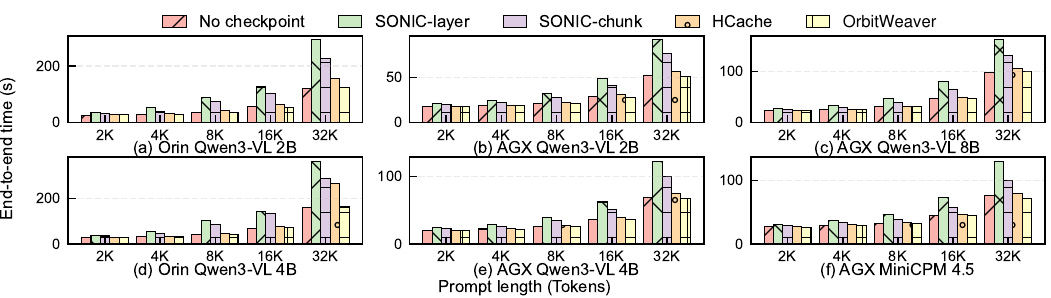}
    \Description{Six bar charts compare uninterrupted inference time across checkpointing baselines and OrbitWeaver at different devices, models, and prompt lengths.}
    \caption{Inference latency under different setups when no shutdowns occur.}
    \label{fig:checkpoint_overhead}
\end{figure*}

\para{OrbitWeaver recovers quickly after resets.}
We use a 32K-token request and create checkpoints at 20\%, 40\%, 60\%, and 80\% of prefill progress, with each request generating 64 tokens during decoding.
\Fref{fig:recovery_latency} reports cold-cache recovery latency at these progress points.
Across all tested devices, models, and progress points, \sysname achieves recovery speedups of $1.22\times$--$6.97\times$ over the baselines.
For example, on AGX with Qwen3-VL 4B at 20\% progress, SONIC-layer, SONIC-chunk, and HCache take $6.85\times$, $6.36\times$, and $4.40\times$ as long as \sysname, respectively.
Even at 80\% progress with MiniCPM-V 4.5, they take $2.95\times$, $2.56\times$, and $1.50\times$ as long, respectively.
This improvement comes from \sysname's dependency-aware recovery: it prioritizes the layer weights, prefix KV cache, and hidden states that unlock useful computation, defers other reads until their consumer blocks are ready, and overlaps SSD reads with dependency-ready computation.
By contrast, fixed chunk-first or layer-first recovery exposes more I/O before computation can resume.

\subsection{Checkpoint overhead}
\label{sec:evaluation:checkpoint_overhead}

\para{OrbitWeaver runs with low normal-path overhead.}
\Fref{fig:checkpoint_overhead} reports uninterrupted inference latency across the tested devices and models with prompt lengths from 2K to 32K tokens, with each request generating 64 tokens during decoding.
Part of each prompt is occupied by image tokens, and the rest is filled with text tokens.
Across all settings, \sysname remains within 5\% of the no-checkpoint execution time while preserving recoverable progress.
Compared with SONIC-layer, SONIC-chunk, and HCache, \sysname achieves speedups of up to $2.63\times$, $2.16\times$, and $1.64\times$, respectively; the corresponding geometric-mean speedups are $1.62\times$, $1.38\times$, and $1.08\times$.
This low normal-path overhead comes from \sysname's block-wise checkpointing, which avoids persisting full-sequence hidden states after every layer, and its storage layout, which writes the states of each completed block as a contiguous record.
Together, these designs reduce device-to-host transfers and fragmented SSD writes during inference.

\subsection{Disk wear}
\label{sec:evaluation:disk_wear}

\begin{figure*}[t]
    \centering
    \begin{minipage}[t]{0.65\textwidth}
        \centering
        \includegraphics[width=\linewidth]{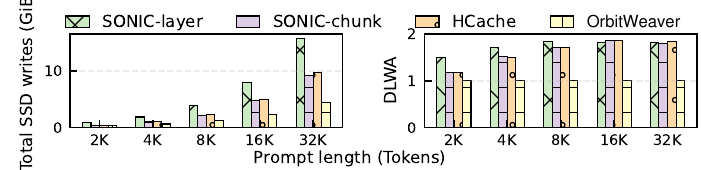}
        \Description{Bar charts compare total SSD writes and device-level write amplification across checkpointing methods.}
        \caption{Total SSD writes and device-level write amplification (DLWA).}
        \label{fig:ssd_write_amount}
    \end{minipage}
    \hfill
    \begin{minipage}[t]{0.32\textwidth}
        \centering
        \includegraphics[width=\linewidth]{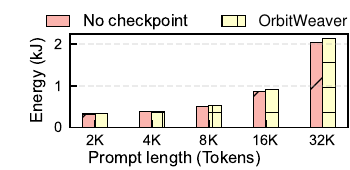}
        \Description{A bar chart compares energy consumption across checkpointing methods.}
        \caption{Energy consumption.}
        \label{fig:energy_consumption}
    \end{minipage}
\end{figure*}

\para{OrbitWeaver reduces SSD writes and write amplification.}
\Fref{fig:ssd_write_amount} reports total SSD writes and DLWA for Qwen3-VL 2B on Orin, with prompt lengths of 2K--32K tokens and 64 generated tokens per request.
Across these lengths, SONIC-layer, SONIC-chunk, and HCache write $2.33\times$--$3.56\times$, $1.08\times$--$2.13\times$, and $1.10\times$--$2.20\times$ as much data as \sysname, respectively.
Compared with SONIC-layer, SONIC-chunk, and HCache, \sysname reduces DLWA by 33.6--45.4\%, 14.3--46.4\%, and 15.0--46.2\%, respectively.
These reductions come from \sysname's phase-aware checkpointing, which avoids persisting redundant intermediate states, and its storage layout, which batches checkpoint states into aligned, contiguous records and produces a mostly sequential write stream.
Together, these designs minimize SSD wear under repeated checkpointing.

\subsection{Energy consumption}
\label{sec:evaluation:energy_consumption}

\para{OrbitWeaver incurs low energy overhead.}
\Fref{fig:energy_consumption} compares uninterrupted Qwen3-VL 2B inference on Orin with and without \sysname checkpointing, measuring energy for 2K--32K prompt tokens and 64 generated tokens per request.
Across all tested prompt lengths, \sysname consumes only 3.75\%--4.74\% more energy than the no-checkpoint baseline.
This overhead comes from persisting recoverable states during normal inference and is consistent with the low runtime overhead in \Fref{fig:checkpoint_overhead}.
The no-checkpoint baseline provides a lower bound for an uninterrupted run but loses all in-memory progress after a reset.
\sysname therefore pays a small persistence cost to avoid repeating completed VLM computation after interruptions.

\subsection{Case Study: Real Deployment on \satelliteone and \satellitetwo}
\label{sec:evaluation:case_study}

\begin{figure}
    \centering
    \includegraphics{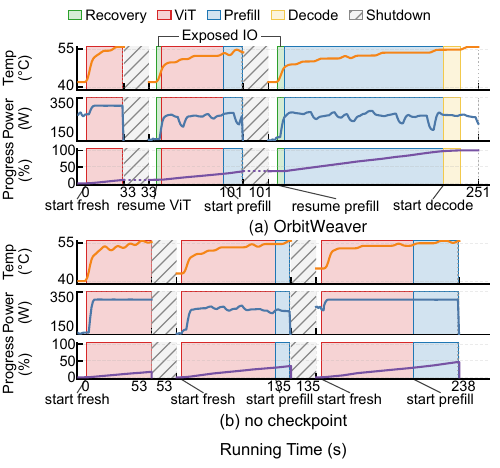}
    \Description{Two payload enclosure temperature and power traces compare OrbitWeaver and no-checkpoint execution across three interrupted execution windows. OrbitWeaver resumes after shutdowns and reaches decoding, while no checkpoint restarts from the beginning after each shutdown and fails to complete.}
    \caption{An example in-orbit VLM execution across interrupted windows on \satelliteone. Temperature refers to the payload enclosure temperature.}
    \label{fig:real_trace}
\end{figure}

\para{OrbitWeaver sustains VLM execution across repeated in-orbit interruptions.}
We deploy \sysname on \satelliteone and \satellitetwo for 72 hours and execute 25 VLM requests on each satellite.
Each request runs Qwen3-VL 2B on remote-sensing images with a 32K-token input, with each request generating 64 tokens during decoding.
We impose a 20-minute timeout on each request.
On \satelliteone, 21 requests experience 47 bus-enforced shutdowns; on \satellitetwo, 20 requests experience 44 bus-enforced shutdowns.
Across both deployments, 41 of the 50 requests are interrupted, and \sysname completes all 50 requests despite a total of 91 bus-enforced shutdowns.

\para{OrbitWeaver preserves progress across bus-enforced shutdowns.}
\Fref{fig:real_trace} shows one execution on \satelliteone that spans three execution windows and two bus-enforced shutdowns.
After each shutdown, \sysname restores the latest committed checkpoint during recovery (green regions) and resumes the unfinished request, completing it in the third window.
For comparison, \Fref{fig:real_trace}(b) shows a separate no-checkpoint execution under repeated bus-enforced shutdowns.
Because the baseline loses all in-memory VLM states after each shutdown, its inference progress (purple line) resets to zero.
It therefore repeats previously completed work and remains in prefill after the third window.
The aggregate deployment demonstrates that \sysname repeatedly recovers across two satellites, while this trace illustrates how preserving progress enables completion across interrupted execution windows.

%% file: sections/discussion.tex
\section{Discussion}
\para{Generalizability.}
\sysname primarily targets orbital edge computing~\cite{asplos2020oec} to reduce reliance on ground-based processing.
Starcloud~\cite{feilden2024space} and Project Suncatcher~\cite{google2025suncatcher} envision dedicated space-based clouds.
Better power and thermal provisioning may reduce interruptions, but inference progress must still survive power cycling, resource contention, and unexpected resets.
\sysname's checkpointing and recovery principles apply to these platforms, complementing hardware co-design by letting interrupted requests resume rather than restart.

%% file: sections/related_work.tex
\section{Related Work}
\label{sec:related}

\parabullet{Satellite Computing}
Satellite computing systems use onboard perception and decision-making to reduce downlink demand and response latency~\cite{asplos2020oec,giuffrida2022phi1,asplos2023kodan,asplos2024eagleeye,infocom2024phoenix,asplos2025earthplus,grace,wang2025viso,leyvamayorga2023satellite_edge_computing,zhai2024seco,zhang2024targetfuse,tao2024serval,furutanpey2025fool,yang2024adaptive_compression_satellite,chen2025slice,zhang2025satcooper}.
Phi-Sat-1 and Kodan reduce downlink demand through onboard filtering~\cite{giuffrida2022phi1,asplos2023kodan}.
EagleEye couples onboard target detection with follow-up observation scheduling across a constellation~\cite{asplos2024eagleeye}.
Grace and VISO adapt VLMs to remote sensing and on-orbit execution~\cite{grace,wang2025viso}.
On-satellite COTS measurements show that thermal and energy constraints strongly shape onboard performance~\cite{mobicom2024cots}.
\sysname instead preserves VLM progress across resets so long-running requests complete despite intermittent satellite execution.

\parabullet{Intermittent Computing}
Intermittent computing systems preserve progress across repeated power or state loss~\cite{mementos_asplos2011,dino,VanDerWoude2016Ratchet,colin2016chain,maeng2017alpaca,maeng2018chinchilla,ganesan2019whatsnext,surbatovich2020formal,kortbeek2022wario,surbatovich2023type,wu2024intos,mohapatra2025datacache}.
Mementos uses checkpoints on energy-harvesting devices~\cite{mementos_asplos2011}.
DINO introduces task boundaries for consistent intermittent execution~\cite{dino}.
Chain provides tasks and channels for intermittent execution~\cite{colin2016chain}.
Alpaca avoids whole-program checkpoints through task-based privatization~\cite{maeng2017alpaca}.
SONIC and Zygarde apply intermittent execution to embedded DNN inference~\cite{asplos2019sonic,islam2020zygarde}.
\sysname extends checkpointing and task-based execution to interruptions from diverse causes, windows of tens of seconds to minutes, and multi-GiB KV and activation states persisted to SSD/eMMC.
The challenges shift to balancing checkpoint size against preserved progress and scheduling recovery I/O with computation.

\parabullet{Efficient LLM and VLM Inference}
Modern VLMs combine visual encoders with language models for flexible visual reasoning and instruction following~\cite{radford2021clip,alayrac2022flamingo,li2023blip2,liu2023llava,dai2023instructblip,peng2024kosmos2,chen2024internvl,wang2024cogvlm}.
LLM serving engines, also used as VLM backends, optimize continuous execution through batching, scheduling, memory offloading, attention kernels, structured generation, prefix/KV cache reuse, and phase disaggregation~\cite{yu2022orca,dao2022flashattention,dao2024flashattention2,sheng2023flexgen,kwon2023vllm,zheng2024sglang,agrawal2024sarathi,zhong2024distserve,patel2024splitwise,miao2024specinfer,jin2023s3}.
Long-context inference systems further reduce KV cache memory footprint or movement through eviction, quantization, compression, streaming, offloading, prefetching, or cache-centric disaggregation~\cite{zhang2023h2o,liu2023scissorhands,xiao2024streamingllm,liu2024kivi,hooper2024kvquant,li2024snapkv,lee2024infinigen,liu2024cachegen,gao2025hcache,qin2025mooncake}.
Unlike these systems, OrbitWeaver targets intermittent on-orbit inference, where volatile activations and KV caches are lost after resets unless they are made durable and recovered efficiently.

%% file: sections/conclusion.tex
\section{Conclusion}
\label{sec:conclusion}

This paper presents \sysname, a runtime for efficient and reliable on-satellite VLM inference through efficient checkpointing and dynamic recovery.
Evaluation demonstrates up to $4.07\times$ end-to-end speedup with less than 5\% runtime overhead, while in-orbit deployment confirms sustained progress across bus-enforced shutdowns.

%% file: refs.bib
@inproceedings{iclr2021vit,
  author    = {Dosovitskiy, Alexey and Beyer, Lucas and Kolesnikov, Alexander and Weissenborn, Dirk and Zhai, Xiaohua and Unterthiner, Thomas and Dehghani, Mostafa and Minderer, Matthias and Heigold, Georg and Gelly, Sylvain and Uszkoreit, Jakob and Houlsby, Neil},
  title     = {An Image is Worth 16x16 Words: Transformers for Image Recognition at Scale},
  booktitle = {International Conference on Learning Representations},
  publisher = {OpenReview.net},
  address   = {Online},
  year      = {2021},
  url       = {https://openreview.net/forum?id=YicbFdNTTy}
}

@techreport{nvme_nvm_command_set_1_0c,
  author      = {{NVM Express, Inc.}},
  title       = {{NVM Express NVM Command Set Specification, Revision 1.0c}},
  institution = {{NVM Express, Inc.}},
  type        = {Specification},
  number      = {Revision 1.0c},
  year        = {2022},
  month       = oct,
  url         = {https://nvmexpress.org/wp-content/uploads/NVM-Express-NVM-Command-Set-Specification-1.0c-2022.10.03-Ratified.pdf}
}

@inproceedings{li2023blip2,
  author    = {Li, Junnan and Li, Dongxu and Savarese, Silvio and Hoi, Steven},
  title     = {{BLIP}-2: Bootstrapping Language-Image Pre-training with Frozen Image Encoders and Large Language Models},
  booktitle = {Proceedings of the 40th International Conference on Machine Learning},
  year      = {2023},
  pages     = {19730--19742},
  volume    = {202},
  series    = {Proceedings of Machine Learning Research},
  publisher = {PMLR},
  address   = {Honolulu, HI, USA},
  url       = {https://proceedings.mlr.press/v202/li23q.html}
}

@inproceedings{liu2023llava,
  author    = {Liu, Haotian and Li, Chunyuan and Wu, Qingyang and Lee, Yong Jae},
  title     = {Visual Instruction Tuning},
  booktitle = {Advances in Neural Information Processing Systems},
  year      = {2023},
  volume    = {36},
  pages     = {34892--34916},
  publisher = {Curran Associates, Inc.},
  address   = {New Orleans, LA, USA},
  url       = {https://arxiv.org/abs/2304.08485}
}

@inproceedings{dino,
author = {Lucia, Brandon and Ransford, Benjamin},
title = {A simpler, safer programming and execution model for intermittent systems},
year = {2015},
isbn = {9781450334686},
publisher = {Association for Computing Machinery},
address = {New York, NY, USA},
url = {https://doi.org/10.1145/2737924.2737978},
doi = {10.1145/2737924.2737978},
booktitle = {Proceedings of the 36th ACM SIGPLAN Conference on Programming Language Design and Implementation},
pages = {575--585},
numpages = {11},
location = {Portland, OR, USA},
}

@techreport{nasa_cii_hosted_payload_guidelines_2011,
  author      = {{Common Instrument Interface Project}},
  title       = {Hosted Payload Guidelines Document},
  institution = {Earth System Science Pathfinder Program Office, NASA Langley Research Center},
  number      = {CII-CI-0001},
  version     = {Baseline},
  year        = {2011},
  month       = nov,
  day         = {14},
  url         = {https://essp.larc.nasa.gov/EV-I/pdf_files/CII_Hosted_Payload_Guidelines_Final.pdf},
  urldate     = {2026-05-15}
}

@online{nasa_basics_power_distribution,
  author       = {{National Aeronautics and Space Administration}},
  title        = {Basics of Space Flight: Chapter 11---Onboard Systems, Electrical Power Distribution},
  organization = {NASA Science},
  year         = {2025},
  url          = {https://science.nasa.gov/learn/basics-of-space-flight/chapter11-3/},
  urldate      = {2026-09-23}
}

@inproceedings{mementos_asplos2011,
  author    = {Ransford, Benjamin and Sorber, Jacob and Fu, Kevin},
  title     = {{Mementos}: System Support for Long-Running Computation on {RFID}-Scale Devices},
  booktitle = {Proceedings of the 16th International Conference on Architectural Support for Programming Languages and Operating Systems},
  year      = {2011},
  pages     = {159--170},
  publisher = {Association for Computing Machinery},
  address   = {New York, NY, USA},
  location  = {Newport Beach, CA, USA},
  doi       = {10.1145/1950365.1950386},
  url       = {https://doi.org/10.1145/1950365.1950386}
}

@inproceedings{VanDerWoude2016Ratchet,
  author    = {Van Der Woude, Joel and Hicks, Matthew},
  title     = {Intermittent Computation without Hardware Support or Programmer Intervention},
  booktitle = {12th USENIX Symposium on Operating Systems Design and Implementation (OSDI 16)},
  year      = {2016},
  month     = nov,
  pages     = {17--32},
  publisher = {USENIX Association},
  address   = {Berkeley, CA, USA},
}

@inproceedings{asplos2019sonic,
  author    = {Gobieski, Graham and Lucia, Brandon and Beckmann, Nathan},
  title     = {Intelligence Beyond the Edge: Inference on Intermittent Embedded Systems},
  booktitle = {Proceedings of the Twenty-Fourth International Conference on Architectural Support for Programming Languages and Operating Systems},
  year      = {2019},
  pages     = {199--213},
  publisher = {Association for Computing Machinery},
  address   = {New York, NY, USA},
  location  = {Providence, RI, USA},
  doi       = {10.1145/3297858.3304011},
  url       = {https://doi.org/10.1145/3297858.3304011}
}

@inproceedings{asplos2024eagleeye,
  author    = {Cheng, Zhuo and Denby, Bradley and McCleary, Kyle and Lucia, Brandon},
  title     = {{EagleEye}: Nanosatellite Constellation Design for High-Coverage, High-Resolution Sensing},
  booktitle = {Proceedings of the 29th ACM International Conference on Architectural Support for Programming Languages and Operating Systems, Volume 1},
  year      = {2024},
  pages     = {117--132},
  publisher = {Association for Computing Machinery},
  address   = {New York, NY, USA},
  location  = {La Jolla, CA, USA},
  doi       = {10.1145/3617232.3624851},
  url       = {https://doi.org/10.1145/3617232.3624851}
}

@inproceedings{asplos2025earthplus,
  author    = {Du, Kuntai and Cheng, Yihua and Olsen, Peder and Noghabi, Shadi and Jiang, Junchen},
  title     = {{Earth+}: On-Board Satellite Imagery Compression Leveraging Historical Earth Observations},
  booktitle = {Proceedings of the 30th ACM International Conference on Architectural Support for Programming Languages and Operating Systems, Volume 1},
  year      = {2025},
  pages     = {361--376},
  publisher = {Association for Computing Machinery},
  address   = {New York, NY, USA},
  location  = {Rotterdam, Netherlands},
  doi       = {10.1145/3669940.3707222},
  url       = {https://doi.org/10.1145/3669940.3707222}
}

@inproceedings{mobicom2024cots,
  author    = {Xing, Ruolin and Xu, Mengwei and Zhou, Ao and Li, Qing and Zhang, Yiran and Qian, Feng and Wang, Shangguang},
  title     = {Deciphering the Enigma of Satellite Computing with {COTS} Devices: Measurement and Analysis},
  booktitle = {Proceedings of the 30th Annual International Conference on Mobile Computing and Networking},
  year      = {2024},
  pages     = {420--435},
  publisher = {Association for Computing Machinery},
  address   = {New York, NY, USA},
  location  = {Washington D.C., DC, USA},
  doi       = {10.1145/3636534.3649371},
  url       = {https://doi.org/10.1145/3636534.3649371}
}

@inproceedings{infocom2024phoenix,
  author    = {Liu, Weisen and Lai, Zeqi and Wu, Qian and Li, Hewu and Zhang, Qi and Li, Zonglun and Li, Yuanjie and Liu, Jun},
  title     = {In-Orbit Processing or Not? Sunlight-Aware Task Scheduling for Energy-Efficient Space Edge Computing Networks},
  booktitle = {IEEE INFOCOM 2024 - IEEE Conference on Computer Communications},
  year      = {2024},
  pages     = {881--890},
  publisher = {IEEE},
  address   = {Piscataway, NJ, USA},
  doi       = {10.1109/INFOCOM52122.2024.10621268},
  url       = {https://doi.org/10.1109/INFOCOM52122.2024.10621268}
}

@inproceedings{asplos2020oec,
  author    = {Denby, Bradley and Lucia, Brandon},
  title     = {Orbital Edge Computing: Nanosatellite Constellations as a New Class of Computer System},
  booktitle = {Proceedings of the Twenty-Fifth International Conference on Architectural Support for Programming Languages and Operating Systems},
  year      = {2020},
  pages     = {939--954},
  publisher = {Association for Computing Machinery},
  address   = {New York, NY, USA},
  location  = {Lausanne, Switzerland},
  doi       = {10.1145/3373376.3378473},
  url       = {https://doi.org/10.1145/3373376.3378473}
}

@inproceedings{asplos2023kodan,
  author    = {Denby, Bradley and Chintalapudi, Krishna and Chandra, Ranveer and Lucia, Brandon and Noghabi, Shadi},
  title     = {{Kodan}: Addressing the Computational Bottleneck in Space},
  booktitle = {Proceedings of the 28th ACM International Conference on Architectural Support for Programming Languages and Operating Systems, Volume 3},
  year      = {2023},
  pages     = {392--403},
  publisher = {Association for Computing Machinery},
  address   = {New York, NY, USA},
  location  = {Vancouver, BC, Canada},
  doi       = {10.1145/3582016.3582043},
  url       = {https://doi.org/10.1145/3582016.3582043}
}

@article{giuffrida2022phi1,
  author  = {Giuffrida, Gianluca and Fanucci, Luca and Meoni, Gabriele and Batic, Matej and Buckley, Leonie and Dunne, Aubrey and van Dijk, Chris and Esposito, Marco and Hefele, John and Vercruyssen, Nathan and Furano, Gianluca and Pastena, Massimiliano and Aschbacher, Josef},
  title   = {The $\Phi$-{Sat}-1 Mission: The First On-Board Deep Neural Network Demonstrator for Satellite Earth Observation},
  journal = {IEEE Transactions on Geoscience and Remote Sensing},
  year    = {2022},
  volume  = {60},
  pages   = {1--14},
  doi     = {10.1109/TGRS.2021.3125567},
  url     = {https://doi.org/10.1109/TGRS.2021.3125567}
}

@article{rs2021fpgacnn,
  author  = {Rapuano, Emilio and Meoni, Gabriele and Pacini, Tommaso and Dinelli, Gianmarco and Furano, Gianluca and Giuffrida, Gianluca and Fanucci, Luca},
  title   = {An {FPGA}-Based Hardware Accelerator for {CNNs} Inference on Board Satellites: Benchmarking with {Myriad} 2-Based Solution for the {CloudScout} Case Study},
  journal = {Remote Sensing},
  year    = {2021},
  volume  = {13},
  number  = {8},
  pages   = {1518},
  doi     = {10.3390/rs13081518},
  url     = {https://doi.org/10.3390/rs13081518}
}

@inproceedings{grace,
author = {Zhang, Yuxin and Yang, Jiahao and Chen, Zhe and Zhu, Wenjun and Zhao, Jin and Gao, Yue},
title = {A Satellite-Ground Synergistic Large Vision-Language Model System for Earth Observation},
year = {2025},
isbn = {9798400720352},
publisher = {Association for Computing Machinery},
address = {New York, NY, USA},
url = {https://doi.org/10.1145/3746027.3755299},
doi = {10.1145/3746027.3755299},
booktitle = {Proceedings of the 33rd ACM International Conference on Multimedia},
pages = {11825--11833},
numpages = {9},
}

@inproceedings{wang2025viso,
  author = {Wang, Meiqi and Qiu, Han},
  title = {VISO: Accelerating In-orbit Object Detection with Language-Guided Mask Learning and Sparse Inference},
  booktitle = {Proceedings of the IEEE/CVF International Conference on Computer Vision},
  series = {ICCV '25},
  pages = {23300--23310},
  year = {2025},
  publisher = {IEEE Computer Society},
  address = {Los Alamitos, CA, USA}
}

@inproceedings{colin2016chain,
  author = {Colin, Alexei and Lucia, Brandon},
  title = {Chain: Tasks and Channels for Reliable Intermittent Programs},
  booktitle = {Proceedings of the 2016 ACM SIGPLAN International Conference on Object-Oriented Programming, Systems, Languages, and Applications},
  series = {OOPSLA '16},
  pages = {514--530},
  year = {2016},
  publisher = {Association for Computing Machinery},
  address = {New York, NY, USA},
  doi = {10.1145/2983990.2983995}
}

@article{maeng2017alpaca,
  author = {Maeng, Kiwan and Colin, Alexei and Lucia, Brandon},
  title = {Alpaca: Intermittent Execution without Checkpoints},
  journal = {Proceedings of the ACM on Programming Languages},
  volume = {1},
  number = {OOPSLA},
  articleno = {96},
  pages = {1--30},
  numpages = {30},
  year = {2017},
  doi = {10.1145/3133920}
}

@inproceedings{maeng2018chinchilla,
  author = {Maeng, Kiwan and Lucia, Brandon},
  title = {Adaptive Dynamic Checkpointing for Safe Efficient Intermittent Computing},
  booktitle = {13th USENIX Symposium on Operating Systems Design and Implementation},
  series = {OSDI '18},
  pages = {129--144},
  year = {2018},
  publisher = {USENIX Association},
  address = {Berkeley, CA, USA}
}

@article{islam2020zygarde,
  author = {Islam, Bashima and Nirjon, Shahriar},
  title = {Zygarde: Time-Sensitive On-Device Deep Inference and Adaptation on Intermittently-Powered Systems},
  journal = {Proceedings of the ACM on Interactive, Mobile, Wearable and Ubiquitous Technologies},
  volume = {4},
  number = {3},
  articleno = {101},
  pages = {1--29},
  numpages = {29},
  year = {2020},
  doi = {10.1145/3411808}
}

@inproceedings{radford2021clip,
  author = {Radford, Alec and Kim, Jong Wook and Hallacy, Chris and Ramesh, Aditya and Goh, Gabriel and Agarwal, Sandhini and Sastry, Girish and Askell, Amanda and Mishkin, Pamela and Clark, Jack and Krueger, Gretchen and Sutskever, Ilya},
  title = {Learning Transferable Visual Models From Natural Language Supervision},
  booktitle = {Proceedings of the 38th International Conference on Machine Learning},
  series = {Proceedings of Machine Learning Research},
  volume = {139},
  pages = {8748--8763},
  year = {2021},
  publisher = {PMLR},
  address = {Online}
}

@inproceedings{alayrac2022flamingo,
  author = {Alayrac, Jean-Baptiste and Donahue, Jeff and Luc, Pauline and Miech, Antoine and Barr, Iain and Hasson, Yana and Lenc, Karel and Mensch, Arthur and Millican, Katie and Reynolds, Malcolm and Ring, Roman and Rutherford, Eliza and Cabi, Serkan and Han, Tengda and Gong, Zhitao and Samangooei, Sina and Monteiro, Marianne and Menick, Jacob and Borgeaud, Sebastian and Brock, Andrew and Nematzadeh, Aida and Sharifzadeh, Sahand and Binkowski, Mikolaj and Barreira, Ricardo and Vinyals, Oriol and Zisserman, Andrew and Simonyan, Karen},
  title = {Flamingo: A Visual Language Model for Few-Shot Learning},
  booktitle = {Advances in Neural Information Processing Systems},
  volume = {35},
  pages = {23716--23736},
  publisher = {Curran Associates, Inc.},
  address = {Red Hook, NY, USA},
  year = {2022}
}

@inproceedings{yu2022orca,
  author = {Yu, Gyeong-In and Jeong, Joo Seong and Kim, Geon-Woo and Kim, Soojeong and Chun, Byung-Gon},
  title = {Orca: A Distributed Serving System for Transformer-Based Generative Models},
  booktitle = {16th USENIX Symposium on Operating Systems Design and Implementation},
  series = {OSDI '22},
  pages = {521--538},
  year = {2022},
  publisher = {USENIX Association},
  address = {Berkeley, CA, USA}
}

@inproceedings{kwon2023vllm,
  author = {Kwon, Woosuk and Li, Zhuohan and Zhuang, Siyuan and Sheng, Ying and Zheng, Lianmin and Yu, Cody Hao and Gonzalez, Joseph E. and Zhang, Hao and Stoica, Ion},
  title = {Efficient Memory Management for Large Language Model Serving with {PagedAttention}},
  booktitle = {Proceedings of the 29th Symposium on Operating Systems Principles},
  series = {SOSP '23},
  pages = {611--626},
  year = {2023},
  publisher = {Association for Computing Machinery},
  address = {New York, NY, USA},
  doi = {10.1145/3600006.3613165}
}

@inproceedings{zheng2024sglang,
  author = {Zheng, Lianmin and Yin, Liangsheng and Xie, Zhiqiang and Sun, Chuyue and Huang, Jeff and Yu, Cody Hao and Cao, Shiyi and Kozyrakis, Christos and Stoica, Ion and Gonzalez, Joseph E. and Barrett, Clark W. and Sheng, Ying},
  title = {SGLang: Efficient Execution of Structured Language Model Programs},
  booktitle = {Advances in Neural Information Processing Systems},
  volume = {37},
  pages = {62557--62583},
  year = {2024},
  publisher = {Curran Associates, Inc.},
  address = {Red Hook, NY, USA}
}

@inproceedings{agrawal2024sarathi,
  author = {Agrawal, Amey and Kedia, Nitin and Panwar, Ashish and Mohan, Jayashree and Kwatra, Nipun and Gulavani, Bhargav and Tumanov, Alexey and Ramjee, Ramachandran},
  title = {Taming Throughput-Latency Tradeoff in LLM Inference with Sarathi-Serve},
  booktitle = {18th USENIX Symposium on Operating Systems Design and Implementation},
  series = {OSDI '24},
  pages = {117--134},
  year = {2024},
  publisher = {USENIX Association},
  address = {Berkeley, CA, USA}
}

@inproceedings{zhong2024distserve,
  author = {Zhong, Yinmin and Liu, Shengyu and Chen, Junda and Hu, Jianbo and Zhu, Yibo and Liu, Xuanzhe and Jin, Xin and Zhang, Hao},
  title = {DistServe: Disaggregating Prefill and Decoding for Goodput-optimized Large Language Model Serving},
  booktitle = {18th USENIX Symposium on Operating Systems Design and Implementation},
  series = {OSDI '24},
  pages = {193--210},
  year = {2024},
  publisher = {USENIX Association},
  address = {Berkeley, CA, USA}
}

@inproceedings{dao2022flashattention,
  author = {Dao, Tri and Fu, Daniel Y. and Ermon, Stefano and Rudra, Atri and R{\'e}, Christopher},
  title = {FlashAttention: Fast and Memory-Efficient Exact Attention with IO-Awareness},
  booktitle = {Advances in Neural Information Processing Systems},
  volume = {35},
  pages = {16344--16359},
  year = {2022},
  publisher = {Curran Associates, Inc.},
  address = {Red Hook, NY, USA}
}

@inproceedings{sheng2023flexgen,
  author = {Sheng, Ying and Zheng, Lianmin and Yuan, Binhang and Li, Zhuohan and Ryabinin, Max and Fu, Daniel Y. and Xie, Zhiqiang and Chen, Beidi and Barrett, Clark W. and Gonzalez, Joseph E. and Liang, Percy and R{\'e}, Christopher and Stoica, Ion and Zhang, Ce},
  title = {FlexGen: High-Throughput Generative Inference of Large Language Models with a Single GPU},
  booktitle = {Proceedings of the 40th International Conference on Machine Learning},
  series = {Proceedings of Machine Learning Research},
  volume = {202},
  pages = {31094--31116},
  year = {2023},
  publisher = {PMLR},
  address = {Honolulu, HI, USA}
}

@inproceedings{liu2024kivi,
  author = {Liu, Zirui and Yuan, Jiayi and Jin, Hongye and Zhong, Shaochen and Xu, Zhaozhuo and Braverman, Vladimir and Chen, Beidi and Hu, Xia},
  title = {KIVI: A Tuning-Free Asymmetric 2bit Quantization for KV Cache},
  booktitle = {Proceedings of the 41st International Conference on Machine Learning},
  series = {Proceedings of Machine Learning Research},
  volume = {235},
  pages = {32332--32344},
  year = {2024},
  publisher = {PMLR},
  address = {Vienna, Austria}
}

@misc{zandieh2025turboquant,
  author        = {Zandieh, Amir and Daliri, Majid and Hadian, Majid and Mirrokni, Vahab},
  title         = {{TurboQuant}: Online Vector Quantization with Near-optimal Distortion Rate},
  year          = {2025},
  eprint        = {2504.19874},
  archivePrefix = {arXiv},
  primaryClass  = {cs.LG},
  doi           = {10.48550/arXiv.2504.19874},
  url           = {https://arxiv.org/abs/2504.19874}
}

@inproceedings{hooper2024kvquant,
  author = {Hooper, Coleman and Kim, Sehoon and Mohammadzadeh, Hiva and Mahoney, Michael W. and Shao, Yakun Sophia and Keutzer, Kurt and Gholami, Amir},
  title = {KVQuant: Towards 10 Million Context Length LLM Inference with KV Cache Quantization},
  booktitle = {Advances in Neural Information Processing Systems},
  volume = {37},
  pages = {1270--1303},
  year = {2024},
  publisher = {Curran Associates, Inc.},
  address = {Red Hook, NY, USA}
}

@misc{google2025suncatcher,
  title        = {Towards a Future Space-Based, Highly Scalable AI Infrastructure System Design},
  author       = {Ag{\"u}era y Arcas, Blaise and Beals, Travis and Biggs, Maria and Bloom, Jessica V. and Fischbacher, Thomas and Gromov, Konstantin and K{\"o}ster, Urs and Pravahan, Rishiraj and Manyika, James},
  year         = {2025},
  eprint       = {2511.19468},
  archivePrefix = {arXiv},
  primaryClass = {cs.DC},
  url          = {https://arxiv.org/abs/2511.19468}
}

@misc{fcc2026spacex_orbital_data_centers,
  title        = {Space Bureau Accepts for Filing SpaceX's Application for Orbital Data Centers},
  author       = {{Federal Communications Commission}},
  year         = {2026},
  month        = feb,
  day          = {4},
  howpublished = {\url{https://docs.fcc.gov/public/attachments/DA-26-113A1.pdf}},
  note         = {DA 26-113; ICFS File No. SAT-LOA-20260108-00016}
}

@inproceedings{geochat,
  title        = {GeoChat: Grounded Large Vision-Language Model for Remote Sensing},
  author       = {Kuckreja, Kartik and Danish, Muhammad Sohail and Naseer, Muzammal and Das, Abhijit and Khan, Salman and Khan, Fahad Shahbaz},
  booktitle    = {Proceedings of the IEEE/CVF Conference on Computer Vision and Pattern Recognition},
  pages        = {27831--27840},
  year         = {2024},
  publisher    = {IEEE Computer Society},
  address      = {Los Alamitos, CA, USA},
  doi          = {10.1109/CVPR52733.2024.02629}
}

@inproceedings{gao2025hcache,
  title     = {Fast State Restoration in LLM Serving with HCache},
  author    = {Gao, Shiwei and Chen, Youmin and Shu, Jiwu},
  booktitle = {Proceedings of the Twentieth European Conference on Computer Systems},
  series    = {EuroSys '25},
  year      = {2025},
  location  = {Rotterdam, Netherlands},
  publisher = {Association for Computing Machinery},
  address   = {New York, NY, USA},
  numpages  = {16},
  doi       = {10.1145/3689031.3696072},
  isbn      = {979-8-4007-1196-1}
}

@inproceedings{luo2025lrsvqa,
  title     = {When Large Vision-Language Model Meets Large Remote Sensing Imagery: Coarse-to-Fine Text-Guided Token Pruning},
  author    = {Luo, Junwei and Zhang, Yingying and Yang, Xue and Wu, Kang and Zhu, Qi and Liang, Lei and Chen, Jingdong and Li, Yansheng},
  booktitle = {Proceedings of the IEEE/CVF International Conference on Computer Vision (ICCV)},
  pages     = {9206--9217},
  month     = {October},
  year      = {2025},
  publisher = {IEEE Computer Society},
  address   = {Los Alamitos, CA, USA}
}

@online{starcloud2026starcloud1,
  author       = {{Starcloud, Inc.}},
  title        = {{Starcloud-1}},
  organization = {Starcloud},
  year         = {n.d.},
  url          = {https://www.starcloud.com/starcloud-1},
  urldate      = {2026-06-10}
}

@misc{du2025firstonorbitdemo,
  author        = {Du, Andrew and Del Prete, Roberto and Mousist, Alejandro and Manser, Nick and Marre, Fabrice and Barton, Andrew and Seubert, Carl and Meoni, Gabriele and Chin, Tat-Jun},
  title         = {First On-Orbit Demonstration of a Geospatial Foundation Model},
  year          = {2025},
  eprint        = {2512.01181},
  archivePrefix = {arXiv},
  primaryClass  = {cs.LG},
  url           = {https://arxiv.org/abs/2512.01181}
}

@inproceedings{wolf2020transformerslib,
  title = {Transformers: State-of-the-Art Natural Language Processing},
  author = {Wolf, Thomas and Debut, Lysandre and Sanh, Victor and Chaumond, Julien and Delangue, Clement and Moi, Anthony and Cistac, Pierric and Rault, Tim and Louf, Remi and Funtowicz, Morgan and Davison, Joe and Shleifer, Sam and von Platen, Patrick and Ma, Clara and Jernite, Yacine and Plu, Julien and Xu, Canwen and Le Scao, Teven Le and Gugger, Sylvain and Drame, Mariama and Lhoest, Quentin and Rush, Alexander M.},
  booktitle = {Proceedings of the 2020 Conference on Empirical Methods in Natural Language Processing: System Demonstrations},
  year = {2020},
  pages = {38--45},
  publisher = {Association for Computational Linguistics},
  address = {Online},
  doi = {10.18653/v1/2020.emnlp-demos.6}
}

@misc{ding2026longearth,
  title         = {{LongEarth-R1}: Benchmarking and Aligning Vision-Language Models for Long-Horizon Earth Observation Reasoning},
  author        = {Ding, Yupan and Xiao, Jing and Zhang, Zhenyuan and Chen, Chaofeng and Liao, Liang and Xia, Gui-Song and Wang, Mi},
  year          = {2026},
  eprint        = {2608.13344},
  archivePrefix = {arXiv},
  primaryClass  = {cs.CV},
  url           = {https://arxiv.org/abs/2608.13344}
}

@inproceedings{ganesan2019whatsnext,
  author    = {Ganesan, Karthik and San Miguel, Joshua and Enright Jerger, Natalie D.},
  title     = {The What's Next Intermittent Computing Architecture},
  booktitle = {Proceedings of the 25th IEEE International Symposium on High Performance Computer Architecture},
  series    = {HPCA '19},
  year      = {2019},
  pages     = {211--223},
  publisher = {IEEE},
  address   = {Piscataway, NJ, USA},
  doi       = {10.1109/HPCA.2019.00039},
  url       = {https://doi.org/10.1109/HPCA.2019.00039}
}

@article{surbatovich2020formal,
  author    = {Surbatovich, Milijana and Lucia, Brandon and Jia, Limin},
  title     = {Towards a Formal Foundation of Intermittent Computing},
  journal   = {Proceedings of the ACM on Programming Languages},
  volume    = {4},
  number    = {OOPSLA},
  articleno = {163},
  numpages  = {31},
  month     = nov,
  year      = {2020},
  doi       = {10.1145/3428231},
  url       = {https://doi.org/10.1145/3428231}
}

@inproceedings{kortbeek2022wario,
  author    = {Kortbeek, Vito and Ghosh, Souradip and Hester, Josiah D. and Campanoni, Simone and Pawe{\l}czak, Przemys{\l}aw},
  title     = {{WARio}: Efficient Code Generation for Intermittent Computing},
  booktitle = {Proceedings of the 43rd ACM SIGPLAN International Conference on Programming Language Design and Implementation},
  series    = {PLDI '22},
  year      = {2022},
  pages     = {777--791},
  publisher = {Association for Computing Machinery},
  address   = {New York, NY, USA},
  doi       = {10.1145/3519939.3523454},
  url       = {https://doi.org/10.1145/3519939.3523454}
}

@article{surbatovich2023type,
  author    = {Surbatovich, Milijana and Spargo, Naomi and Jia, Limin and Lucia, Brandon},
  title     = {A Type System for Safe Intermittent Computing},
  journal   = {Proceedings of the ACM on Programming Languages},
  volume    = {7},
  number    = {PLDI},
  articleno = {136},
  numpages  = {25},
  month     = jun,
  year      = {2023},
  doi       = {10.1145/3591250},
  url       = {https://doi.org/10.1145/3591250}
}

@inproceedings{wu2024intos,
  author    = {Wu, Yilun and Min, Byounguk and Ismail, Mohannad and Xiong, Wenjie and Jung, Changhee and Lee, Dongyoon},
  title     = {{IntOS}: Persistent Embedded Operating System and Language Support for Multi-threaded Intermittent Computing},
  booktitle = {18th USENIX Symposium on Operating Systems Design and Implementation},
  series    = {OSDI '24},
  year      = {2024},
  pages     = {425--443},
  publisher = {USENIX Association},
  address   = {Santa Clara, CA},
  month     = jul,
  isbn      = {978-1-939133-40-3},
  url       = {https://www.usenix.org/conference/osdi24/presentation/wu-yilun}
}

@inproceedings{mohapatra2025datacache,
  author    = {Mohapatra, Sourav and Kortbeek, Vito and van Eerden, Marco Antonio and Broekhoff, Jochem and Ahmed, Saad and Pawe{\l}czak, Przemys{\l}aw},
  title     = {Data Cache for Intermittent Computing Systems with Non-Volatile Main Memory},
  booktitle = {Proceedings of the 30th ACM International Conference on Architectural Support for Programming Languages and Operating Systems, Volume 2},
  series    = {ASPLOS '25},
  year      = {2025},
  pages     = {227--243},
  publisher = {Association for Computing Machinery},
  address   = {New York, NY, USA},
  doi       = {10.1145/3676641.3715989},
  url       = {https://doi.org/10.1145/3676641.3715989}
}

@inproceedings{dai2023instructblip,
  author    = {Dai, Wenliang and Li, Junnan and Li, Dongxu and Tiong, Anthony Meng Huat and Zhao, Junqi and Wang, Weisheng and Li, Boyang and Fung, Pascale and Hoi, Steven C. H.},
  title     = {{InstructBLIP}: Towards General-purpose Vision-Language Models with Instruction Tuning},
  booktitle = {Advances in Neural Information Processing Systems},
  volume    = {36},
  pages     = {49250--49267},
  year      = {2023},
  publisher = {Curran Associates, Inc.},
  address   = {Red Hook, NY, USA}
}

@inproceedings{peng2024kosmos2,
  author    = {Peng, Zhiliang and Wang, Wenhui and Dong, Li and Hao, Yaru and Huang, Shaohan and Ma, Shuming and Wei, Furu},
  title     = {Grounding Multimodal Large Language Models to the World},
  booktitle = {International Conference on Learning Representations},
  publisher = {OpenReview.net},
  address   = {Vienna, Austria},
  year      = {2024},
  url       = {https://openreview.net/forum?id=lLmqxkfSIw}
}

@inproceedings{chen2024internvl,
  author    = {Chen, Zhe and Wu, Jiannan and Wang, Wenhai and Su, Weijie and Chen, Guo and Xing, Sen and Zhong, Muyan and Zhang, Qinglong and Zhu, Xizhou and Lu, Lewei and Li, Bin and Luo, Ping and Lu, Tong and Qiao, Yu and Dai, Jifeng},
  title     = {{InternVL}: Scaling up Vision Foundation Models and Aligning for Generic Visual-Linguistic Tasks},
  booktitle = {Proceedings of the IEEE/CVF Conference on Computer Vision and Pattern Recognition},
  pages     = {24185--24198},
  year      = {2024},
  publisher = {IEEE Computer Society},
  address   = {Los Alamitos, CA, USA}
}

@inproceedings{wang2024cogvlm,
  author    = {Wang, Weihan and Lv, Qingsong and Yu, Wenmeng and Hong, Wenyi and Qi, Ji and Wang, Yan and Ji, Junhui and Yang, Zhuoyi and Zhao, Lei and Song, Xixuan and Xu, Jiazheng and Chen, Keqin and Xu, Bin and Li, Juanzi and Dong, Yuxiao and Ding, Ming and Tang, Jie},
  title     = {{CogVLM}: Visual Expert for Pretrained Language Models},
  booktitle = {Advances in Neural Information Processing Systems},
  volume    = {37},
  pages     = {121475--121499},
  year      = {2024},
  publisher = {Curran Associates, Inc.},
  address   = {Red Hook, NY, USA}
}

@inproceedings{dao2024flashattention2,
  author    = {Dao, Tri},
  title     = {{FlashAttention-2}: Faster Attention with Better Parallelism and Work Partitioning},
  booktitle = {International Conference on Learning Representations},
  publisher = {OpenReview.net},
  address   = {Vienna, Austria},
  year      = {2024},
  url       = {https://openreview.net/forum?id=mZn2Xyh9Ec}
}

@inproceedings{miao2024specinfer,
  author    = {Miao, Xupeng and Oliaro, Gabriele and Zhang, Zhihao and Cheng, Xinhao and Wang, Zeyu and Zhang, Zhengxin and Wong, Rae Ying Yee and Zhu, Alan and Yang, Lijie and Shi, Xiaoxiang and Shi, Chunan and Chen, Zhuoming and Arfeen, Daiyaan and Abhyankar, Reyna and Jia, Zhihao},
  title     = {{SpecInfer}: Accelerating Large Language Model Serving with Tree-based Speculative Inference and Verification},
  booktitle = {Proceedings of the 29th ACM International Conference on Architectural Support for Programming Languages and Operating Systems},
  series    = {ASPLOS '24},
  year      = {2024},
  pages     = {932--949},
  publisher = {Association for Computing Machinery},
  address   = {New York, NY, USA},
  doi       = {10.1145/3620666.3651335}
}

@inproceedings{patel2024splitwise,
  author    = {Patel, Pratyush and Choukse, Esha and Zhang, Chaojie and Shah, Aashaka and Goiri, {\'I}{\~n}igo and Maleki, Saeed and Bianchini, Ricardo},
  title     = {{Splitwise}: Efficient Generative LLM Inference Using Phase Splitting},
  booktitle = {2024 ACM/IEEE 51st Annual International Symposium on Computer Architecture},
  series    = {ISCA '24},
  year      = {2024},
  pages     = {118--132},
  publisher = {IEEE},
  address   = {Piscataway, NJ, USA},
  doi       = {10.1109/ISCA59077.2024.00019}
}

@inproceedings{jin2023s3,
  author    = {Jin, Yunho and Wu, Chun-Feng and Brooks, David and Wei, Gu-Yeon},
  title     = {{$S^3$}: Increasing GPU Utilization during Generative Inference for Higher Throughput},
  booktitle = {Advances in Neural Information Processing Systems},
  volume    = {36},
  pages     = {18015--18027},
  year      = {2023},
  publisher = {Curran Associates, Inc.},
  address   = {Red Hook, NY, USA}
}

@inproceedings{zhang2023h2o,
  author    = {Zhang, Zhenyu and Sheng, Ying and Zhou, Tianyi and Chen, Tianlong and Zheng, Lianmin and Cai, Ruisi and Song, Zhao and Tian, Yuandong and R{\'e}, Christopher and Barrett, Clark and Wang, Zhangyang and Chen, Beidi},
  title     = {{H2O}: Heavy-Hitter Oracle for Efficient Generative Inference of Large Language Models},
  booktitle = {Advances in Neural Information Processing Systems},
  volume    = {36},
  pages     = {34661--34710},
  year      = {2023},
  publisher = {Curran Associates, Inc.},
  address   = {Red Hook, NY, USA}
}

@inproceedings{liu2023scissorhands,
  author    = {Liu, Zichang and Desai, Aditya and Liao, Fangshuo and Wang, Weitao and Xie, Victor and Xu, Zhaozhuo and Kyrillidis, Anastasios and Shrivastava, Anshumali},
  title     = {Scissorhands: Exploiting the Persistence of Importance Hypothesis for LLM KV Cache Compression at Test Time},
  booktitle = {Advances in Neural Information Processing Systems},
  volume    = {36},
  pages     = {52342--52364},
  year      = {2023},
  publisher = {Curran Associates, Inc.},
  address   = {Red Hook, NY, USA}
}

@inproceedings{xiao2024streamingllm,
  author    = {Xiao, Guangxuan and Tian, Yuandong and Chen, Beidi and Han, Song and Lewis, Mike},
  title     = {Efficient Streaming Language Models with Attention Sinks},
  booktitle = {International Conference on Learning Representations},
  publisher = {OpenReview.net},
  address   = {Vienna, Austria},
  year      = {2024},
  url       = {https://openreview.net/forum?id=NG7sS51zVF}
}

@inproceedings{li2024snapkv,
  author    = {Li, Yuhong and Huang, Yingbing and Yang, Bowen and Venkitesh, Bharat and Locatelli, Acyr and Ye, Hanchen and Cai, Tianle and Lewis, Patrick and Chen, Deming},
  title     = {{SnapKV}: LLM Knows What You are Looking for Before Generation},
  booktitle = {Advances in Neural Information Processing Systems},
  volume    = {37},
  pages     = {22947--22970},
  year      = {2024},
  publisher = {Curran Associates, Inc.},
  address   = {Red Hook, NY, USA}
}

@inproceedings{lee2024infinigen,
  author    = {Lee, Wonbeom and Lee, Jungi and Seo, Junghwan and Sim, Jaewoong},
  title     = {{InfiniGen}: Efficient Generative Inference of Large Language Models with Dynamic KV Cache Management},
  booktitle = {18th USENIX Symposium on Operating Systems Design and Implementation},
  series    = {OSDI '24},
  pages     = {155--172},
  year      = {2024},
  publisher = {USENIX Association},
  address   = {Berkeley, CA, USA}
}

@inproceedings{liu2024cachegen,
  author    = {Liu, Yuhan and Li, Hanchen and Cheng, Yihua and Ray, Siddhant and Huang, Yuyang and Zhang, Qizheng and Du, Kuntai and Yao, Jiayi and Lu, Shan and Ananthanarayanan, Ganesh and Maire, Michael and Hoffmann, Henry and Holtzman, Ari and Jiang, Junchen},
  title     = {{CacheGen}: KV Cache Compression and Streaming for Fast Large Language Model Serving},
  booktitle = {Proceedings of the ACM SIGCOMM 2024 Conference},
  series    = {SIGCOMM '24},
  pages     = {38--56},
  year      = {2024},
  publisher = {Association for Computing Machinery},
  address   = {New York, NY, USA},
  doi       = {10.1145/3651890.3672274}
}

@inproceedings{qin2025mooncake,
  author    = {Qin, Ruoyu and Li, Zheming and He, Weiran and Cui, Jialei and Ren, Feng and Zhang, Mingxing and Wu, Yongwei and Zheng, Weimin and Xu, Xinran},
  title     = {{Mooncake}: Trading More Storage for Less Computation---A {KVCache-centric} Architecture for Serving {LLM} Chatbot},
  booktitle = {23rd USENIX Conference on File and Storage Technologies},
  series    = {FAST '25},
  pages     = {155--170},
  year      = {2025},
  publisher = {USENIX Association},
  address   = {Santa Clara, CA}
}

@article{leyvamayorga2023satellite_edge_computing,
  author  = {Leyva-Mayorga, Israel and Martinez-Gost, Marc and Moretti, Marco and P{\'e}rez-Neira, Ana and V{\'a}zquez, Miguel Angel and Popovski, Petar and Soret, Beatriz},
  title   = {Satellite Edge Computing for Real-Time and Very-High Resolution Earth Observation},
  journal = {IEEE Transactions on Communications},
  year    = {2023},
  volume  = {71},
  number  = {10},
  pages   = {6180--6194},
  doi     = {10.1109/TCOMM.2023.3296584},
  url     = {https://doi.org/10.1109/TCOMM.2023.3296584}
}

@inproceedings{zhai2024seco,
  author    = {Zhai, Zhiwei and Zeng, Liekang and Ouyang, Tao and Yu, Shuai and Huang, Qianyi and Chen, Xu},
  title     = {{SECO}: Multi-Satellite Edge Computing Enabled Wide-Area and Real-Time Earth Observation Missions},
  booktitle = {IEEE INFOCOM 2024 - IEEE Conference on Computer Communications},
  year      = {2024},
  pages     = {2548--2557},
  publisher = {IEEE},
  address   = {Piscataway, NJ, USA},
  doi       = {10.1109/INFOCOM52122.2024.10621270},
  url       = {https://doi.org/10.1109/INFOCOM52122.2024.10621270}
}

@inproceedings{zhang2024targetfuse,
  author    = {Zhang, Qiyang and Yuan, Xin and Xing, Ruolin and Zhang, Yiran and Zheng, Zimu and Ma, Xiao and Xu, Mengwei and Dustdar, Schahram and Wang, Shangguang},
  title     = {Resource-efficient In-orbit Detection of Earth Objects},
  booktitle = {IEEE INFOCOM 2024 - IEEE Conference on Computer Communications},
  year      = {2024},
  pages     = {551--560},
  publisher = {IEEE},
  address   = {Piscataway, NJ, USA},
  doi       = {10.1109/INFOCOM52122.2024.10621328},
  url       = {https://doi.org/10.1109/INFOCOM52122.2024.10621328}
}

@inproceedings{tao2024serval,
  author    = {Tao, Bill and Chabra, Om and Janveja, Ishani and Gupta, Indranil and Vasisht, Deepak},
  title     = {Known Knowns and Unknowns: Near-realtime Earth Observation Via Query Bifurcation in {Serval}},
  booktitle = {21st USENIX Symposium on Networked Systems Design and Implementation},
  series    = {NSDI '24},
  year      = {2024},
  pages     = {809--824},
  publisher = {USENIX Association},
  address   = {Santa Clara, CA, USA},
  url       = {https://www.usenix.org/conference/nsdi24/presentation/tao}
}

@article{furutanpey2025fool,
  author  = {Furutanpey, Alireza and Zhang, Qiyang and Raith, Philipp and Pfandzelter, Tobias and Wang, Shangguang and Dustdar, Schahram},
  title   = {{FOOL}: Addressing the Downlink Bottleneck in Satellite Computing With Neural Feature Compression},
  journal = {IEEE Transactions on Mobile Computing},
  year    = {2025},
  volume  = {24},
  number  = {8},
  pages   = {6747--6764},
  doi     = {10.1109/TMC.2025.3544516},
  url     = {https://doi.org/10.1109/TMC.2025.3544516}
}

@article{yang2024adaptive_compression_satellite,
  author  = {Yang, Chen and Sun, Qibo and Zhang, Qiyang and Lu, Hao and Ardagna, Claudio A. and Wang, Shangguang and Xu, Mengwei},
  title   = {Toward Efficient Satellite Computing Through Adaptive Compression},
  journal = {IEEE Transactions on Services Computing},
  year    = {2024},
  volume  = {17},
  number  = {6},
  pages   = {4411--4422},
  doi     = {10.1109/TSC.2024.3470341},
  url     = {https://doi.org/10.1109/TSC.2024.3470341}
}

@article{chen2025slice,
  author  = {Chen, Yijie and Zhang, Qiyang and Xing, Ruolin and Li, Yuanzhe and Ma, Xiao and Zhang, Yiran and Zhou, Ao and Wang, Shangguang},
  title   = {{SLICE}: Energy-Efficient Satellite-Ground Co-Inference via Layer-Wise Scheduling Optimization},
  journal = {IEEE Transactions on Services Computing},
  year    = {2025},
  volume  = {18},
  number  = {4},
  pages   = {2388--2402},
  doi     = {10.1109/TSC.2025.3577451},
  url     = {https://doi.org/10.1109/TSC.2025.3577451}
}

@article{zhang2025satcooper,
  author  = {Zhang, Qiyang and Wang, Shangguang and Guan, Jinglong and Donta, Praveen Kumar and Ma, Xiao and Zhang, Yiran and Prasad, RangaRao Venkatesha and Dustdar, Schahram and Liu, Xuanzhe},
  title   = {{SatCooper}: Enhancing Cooperative Inference Analytics for Satellite Service via Multi-Exit {DNNs}},
  journal = {IEEE Transactions on Mobile Computing},
  year    = {2025},
  volume  = {24},
  number  = {9},
  pages   = {8314--8328},
  doi     = {10.1109/TMC.2025.3556457},
  url     = {https://doi.org/10.1109/TMC.2025.3556457}
}

@inproceedings{evans2016opssatfdir,
  author    = {Evans, David and Ortega, Manuel and Zeif, Reinhard and Sergert, Tom},
  title     = {{OPS-SAT}: {FDIR} Design on a Mission that Expects Bugs---and Lots of Them},
  booktitle = {SpaceOps 2016 Conference},
  publisher = {American Institute of Aeronautics and Astronautics},
  address   = {Reston, VA, USA},
  year      = {2016},
  doi       = {10.2514/6.2016-2481}
}

@online{hatfield2018dellingr,
  author       = {Hatfield, Miles},
  title        = {{Dellingr}: The Little {CubeSat} That Could},
  organization = {NASA},
  year         = {2018},
  month        = oct,
  day          = {23},
  url          = {https://www.nasa.gov/solar-system/dellingr-the-little-cubesat-that-could/},
  urldate      = {2026-09-24}
}

@article{ali2026autonomousEPS,
  author  = {Ali, Husan and Iqbal, Asim and Alatwi, Aadel Mohammed and Ali, Akhtar and Khan, Bakht Muhammad},
  title   = {Design and experimental validation of an autonomous electrical power system ({EPS}) for {CubeSats}},
  journal = {Scientific Reports},
  volume  = {16},
  number  = {1},
  articleno = {22981},
  numpages = {18},
  year    = {2026},
  doi     = {10.1038/s41598-026-51416-y}
}

@article{fong2011formosat,
  author  = {Fong, C.-J. and Whiteley, D. and Yang, E. and Cook, K. and Chu, V. and Schreiner, B. and Ector, D. and Wilczynski, P. and Liu, T.-Y. and Yen, N.},
  title   = {Space and ground segment performance and lessons learned of the {FORMOSAT-3/COSMIC} mission: four years in orbit},
  journal = {Atmospheric Measurement Techniques},
  volume  = {4},
  number  = {6},
  pages   = {1115--1132},
  year    = {2011},
  doi     = {10.5194/amt-4-1115-2011}
}

@inproceedings{tavakkol2018mqsim,
  author    = {Tavakkol, Arash and G{\'o}mez-Luna, Juan and Sadrosadati, Mohammad and Ghose, Saugata and Mutlu, Onur},
  title     = {{MQSim}: A Framework for Enabling Realistic Studies of Modern Multi-Queue {SSD} Devices},
  booktitle = {16th USENIX Conference on File and Storage Technologies (FAST 18)},
  year      = {2018},
  pages     = {49--66},
  publisher = {USENIX Association},
  address   = {Berkeley, CA, USA}
}

@techreport{feilden2024space,
  author      = {Feilden, Ezra and Oltean, Adi and Johnston, Philip},
  title       = {Why We Should Train {AI} in Space},
  institution = {Lumen Orbit (now Starcloud)},
  type        = {White Paper},
  number      = {v1.03},
  month       = sep,
  year        = {2024},
  url         = {https://starcloudinc.github.io/wp.pdf}
}
